\documentclass[preprint,12pt]{elsarticle}

\usepackage{graphicx}

\usepackage{amssymb}

\usepackage{dcolumn}                                    
\usepackage{multicol}                                   
\usepackage{booktabs}                                   
\usepackage{hhline}

\usepackage{color}
\usepackage{units}
\usepackage{overpic}

\journal{Astroparticle Physics}

\begin{document}

\begin{frontmatter}

\title{A detailed comparison of REAS3 and MGMR simulations for radio emission from EAS}

\author[KITIEKP]{M. Ludwig\corref{cor}}
\author[KITIK]{T. Huege}
\author[KVI]{O. Scholten}
\author[KVI]{K.\,D. de Vries}

\address[KITIK]{Institut f\"ur Experimentelle Kernphysik, Karlsruher Institut f\"ur Technologie - Campus S\"ud, 76128 Karlsruhe, Germany}
\address[KITIK]{Institut f\"ur Kernphysik, Karlsruher Institut f\"ur Technologie - Campus Nord, Postfach 3640, 76021 Karlsruhe, Germany}
\address[KVI]{Kernfysisch Versneller Instituut, Zernikelaan 25, 9747 AA Groningen, The Netherlands}

\cortext[cor]{Corresponding author: Marianne Ludwig $<$marianne.ludwig@kit.edu$>$}

\begin{abstract}
In the last years, several models and simulations calculating the radio emission from cosmic ray air showers have been developed. However, a number of those made conflicting predictions on the pulse shapes and the amplitudes of the radio signal. In the scope of this paper, we discuss a detailed comparison of two independent and complementary theoretical approaches, namely MGMR and REAS3. Furthermore, we study the influence of the underlying air shower models on the predicted pulse shapes and amplitudes and show that remaining discrepancies between MGMR and REAS3 are mostly determined by the air shower models. With this general agreement, a breakthrough in the understanding of the modelling of radio emission from air showers has been achieved.
\end{abstract}


\end{frontmatter}


\section{Introduction}\label{sec:intro}
Over the past years, the interest in the detection of radio emission from extensive air showers (EAS) increased continuously due to its promising potential and the results derived with experiments such as LOPES \cite{Falcke2005, Link2011} and CODALEMA \cite{Ardouin2005, Ravel2010}. To understand the measurements and to study the cosmic rays with the help of radio detection, detailed simulations and a solid theoretical understanding are needed. In the past, the models REAS2 \cite{HuegeUlrichEngel2007a} and MGMR \cite{Sch08} made conflicting predictions on the pulse shape and the radio pulse amplitudes. While REAS2 predicted unipolar pulses, the pulses calculated with MGMR were bipolar and up to a factor of 10 lower (depending on observation bandwidth and shower geometry) than the amplitudes obtained with REAS2. With REAS3, a missing radiation component resulting from the variation of the number of charged particles within the air shower is taken into account. From that time on, REAS3 predicts bipolar pulses \cite{Ludwig2010}. In MGMR, an additional contribution due to the charge excess was implemented \cite{Wer08}. These revisions motivate a detailed comparison between MGMR and REAS3.\\
In the following, an introduction on both models is given to stress that the models are completely independent and complementary.

\subsection{A macroscopic approach -- MGMR}\label{subs:MGMR}
When an UHE cosmic ray penetrates the atmosphere, an extensive air shower (EAS), a cascade of secondary particles, is created. These particles are concentrated in a relatively small plasma cloud that moves with almost the light velocity. In this plasma cloud a macroscopic current is induced, through the Lorentz force, by the magnetic field of the Earth. The reason is that the induced drift velocity, perpendicular to the direction of the initial cosmic ray has opposite directions for the electrons and positrons. The knock-out of electrons from the air molecules by the high energy photons (through Compton scattering) and leptons builds a net excess of electrons that move with the same velocity as the plasma cloud. The moving charge excess and the induced transverse current can be combined in a four-vector current.
This four-vector is roughly proportional to the number of charged particles in the plasma cloud as the induced drift velocity is varying little with height and the electron knockout is a statistical process. As the number of charged particles is varying with height the resulting variation of the four-current will lead to the emission of electromagnetic radiation. As long as the wavelength of the emitted radiation is considerably longer than the size of the charge cloud and the (retarded) time-variation of the particle number is larger than the oscillation time, all charges contribute coherently resulting in very intense radiation. As a result, the frequency spectrum contains information of the important length and time scales of the four-current in the EAS \cite{Sch09}. The polarisation of the emitted radiation offers additional clues on the origin of the coherent radiation \cite{Wer08,dVries10}.\\
The MGMR-model has its basis in the Li\'enard-Wiechert potentials from classical electrodynamics~\cite{Jac-CE} given by,
\begin{equation}
A^{\mu}(t,\vec{x})=
\frac{\mu _0}{4\pi}\int \mathrm{d}^3\vec{r}\; \frac{J^\mu(t',\vec{r},h)}{|{\cal D}|}\Bigr{|}_{t=t'} \;,
\end{equation}
where $t'$ denotes the negative retarded time corresponding to the time the signal was emitted from the shower front and ${\cal D}$ is the retarded distance~\cite{Wer08}. The volume integral takes care of the particle distributions in the shower front. In MGMR the lateral particle distribution is neglected, hence the four-current can be written like,
\begin{equation}
J^\mu (t',\vec{x},h)=J^\mu(t')\delta^{2}(\vec{r})f(h) \;,
\end{equation}
where $\vec{r}=(x,y)$ is the lateral distance with respect to the shower axis at the position $z=-ct'+h$. Where $z$ is the distance to the Earth's surface as measured along the shower path, and $h$ is the longitudinal distance along the shower axis with respect to the shower front. From this is follows that $f(h)$ describes the normalised longitudinal particle distribution in the shower front. In MGMR, this distribution is fitted with a $\Gamma-$probability distribution function given by,
\begin{equation}\label{eq:longiDisplacement}
f(h) = h \cdot e^{-2h/L} \cdot \frac{4}{L^2}
\end{equation}
where $L$ denotes the thickness of the pancake which is set to 3.9\,m in this comparison. The electromagnetic fields are obtained through the standard relation,
\begin{eqnarray}
\vec{E}(\vec{x},t)&=&-\frac{\mathrm{d}}{\mathrm{d}\vec{x}}A^0-\frac{\mathrm{d}}{\mathrm{d}ct}\vec{A} \;,
\end{eqnarray}
for an observer located at $\vec{x}=(x,y,z)$, and where at $t=t'=0$ the shower hits the Earth. The velocity of the shower front is $c\beta$ where we set $\beta=1$.

\subsection{A microscopic approach -- REAS3}\label{subs:REAS3}
Motivated by the geosynchrotron idea \cite{FalckeGorham2003}, the C++ based Monte Carlo code REAS1 was published in 2005. REAS calculates the radio emission from individual electrons and positrons and superposes their radiation at a given observer position. With the updated version, REAS2, it was possible to calculate radio emission from air showers which had been simulated with CORSIKA \cite{Heck1998}. The air shower information derived with CORSIKA is saved in histograms by an interface program COAST. However, it turned out that REAS2 (and all other microscopic models at this time) was missing a radiation component coming from the variation of the number of charged particles. With the latest version REAS3 \cite{Ludwig2010}, these contributions are now taken into account.\\
In REAS3, the tracks of the single particles are described by straight segments joined by kinks. In each kink, an acceleration process is taking place. The radio emission arising from these acceleration processes is calculated using the endpoint formalism \cite{James2011}. With this approach all emission related to the acceleration of charged particles is taken into account automatically as long as the underlying particle movement ist described with sufficient precision. Hence, no assumptions on the mechanism of radio emission need to be applied. Since also, the radiation at the beginning and the end of the tracks is calculated, the radiation due to the variation of number of charges is considered as well as the radiation due to the deflection of the particles in the Earth's magnetic field. To calculate the emission in each kink, the radiation formula derived from the  Li\'enard-Wiechert potentials is used as the starting point. With this, the radiation is:
\begin{eqnarray}
\int \vec{E}(\vec{x},t) \mathrm{d}t & = & \int_{t_1}^{t_2} \frac{e}{c} \left\vert \frac{ \vec{r}\times [(\vec{r}-\vec{\beta})\times
	 \dot{\vec{\beta}}]} {(1-\vec{\beta}\cdot\vec{r})^3 R}\right\vert_{ret} \mathrm{d}t = \vec{F}(t_2) - \vec{F}(t_1) \nonumber \\
	 &= &\frac{e}{cR}\left(\frac{\vec{r}\times (\vec{r}\times \vec{\beta_2})} {(1-\vec{\beta_2}\vec{r})} \right) - 
	 \frac{e}{cR}\left(\frac{\vec{r}\times (\vec{r}\times \vec{\beta_1})} {(1-\vec{\beta_1}\vec{r})} \right)
\end{eqnarray}
where $\vec{\beta_1}$ corresponds to the particle velocity before and $\vec{\beta_2}$ after the kink while $\vec{\beta} = \vec{v}(t)/c$ is given by the particle velocity. Furthermore, $e$ indicates the particle charge, $R(t) = \vert\vec{R}(t)\vert$ denotes the vector between particle and observer position, $\vec{r}(t) = \vec{R}(t)/R(t)$ is the line-of-sight direction between particle and observer, and $\gamma$ is the Lorentz factor of the particle. The index ``ret'' means that the equation needs to be
evaluated in retarded time. \\
For a starting-point of the track, only the term with $\vec{\beta_2}\cdot c$ gives a contribution since the velocity before the kink is zero, and vice versa, at the end of the track, the contribution is given by the term with $\vec{\beta_1}\cdot c$.

\section{Methodology}\label{sec:method}

To make sure that the results of both models are indeed comparable, a set of prototype showers has been defined for the simulations. All of these air showers have a fixed geometry with a specific energy for the primary particle and the observer positions are well-defined. For each prototype shower, a set of CORSIKA showers has been simulated. To avoid shower-to-shower fluctuations, one typical shower was selected, i.e. one which has a shower maximum $X_{\mbox{max}}$ close to the mean $X_{\mbox{max}}$. The chosen hadronic interaction models were QGSJetII.03  \cite{OstapchenkoQGSjetII2006a,OstapchenkoQGSjetII2006b} and UrQmd1.3.1 \cite{Bass1998,Bleicher1999}. The values for the magnetic field have been set to the Argentinian magnetic field at the site of the Pierre Auger Observatory \cite{Abraham2004}, i.e. to a magnetic field of 0.23\,Gauss with an inclination of $-37^\circ$, except for the showers considered in section \ref{subs:magnetic}. The observer positions were all set to 1400\,m above sea level, which corresponds to the site of the Pierre Auger Observatory. To calculate the radio emission from air showers with MGMR, some parameters need to be set which have been determined from Monte Carlo simulations. For this comparison they are listed in table \ref{tab:mgmrParameters} and published in \cite{dVries10}. \\
To perform the MGMR simulations, we used the longitudinal profiles of the CORSIKA simulated air showers while in REAS3 the histogramed information from the same air shower was taken into account. This ensures that a similar parametrisation of the air shower is used for the MGMR simulations and thus allows a direct comparison of the results. The difference between both, however, is the level of detail in which they take the air shower properties into account.
\begin{table}[tb!]
\begin{center}
\begin{minipage}[b]{0.95\textwidth}
{\footnotesize
\caption[Parameters used by MGMR]{Overview of the parameters set in MGMR-v1.6 used for the comparison with REAS3. These parameters have been determined from Monte Carlo simulations.\\} \label{tab:mgmrParameters}
\begin{tabular}{>{\centering\arraybackslash}m{0.4\textwidth}>{\centering\arraybackslash}m{0.25\textwidth}>{\centering\arraybackslash}m{0.25\textwidth}}\hline
\textbf{Parameter (name)} & \textbf{Value} & \textbf{Comment} \\ \hline
drift velocity $v_d$ & $0.025\cdot c$   &  $c$ = speed of light \\
pancake thickness $L$ & 3.9\,m   & assumed as constant \\
fraction of charge excess & 0.23  & assumed as constant \\
\hline
\end{tabular}
}
\end{minipage}
\end{center}
\end{table}
Furthermore in MGMR, the lateral distribution of the particles in the shower pancake is not taken into account. In section \ref{sec:simplified}, the differences in the air shower models are discussed in more detail.\\
Parametrisations which will not be considered in this article are briefly summarised in the following:\\
In REAS3, the fraction of the charge excess with respect to the total number of charges increases with the shower development as given by the longitudinal development in CORSIKA. In MGMR, the ratio is set to 23\% for the whole air shower which is comparable to the realistic net charge distribution averaged over the longitudinal development of the air shower. However, this leads to differences in the amount of charge excess.\\
Furthermore, the velocity of the pancake is set to the speed of light, i.e. $\beta=1$, in the case of MGMR (cf. section \ref{subs:MGMR}). The assumption here is that due to the constant creation of high energy electrons and positrons in the shower front, the macroscopic particle distribution has the speed of light. A direct consequence is that at the back of the pancake a constant flow of low energetic particles will be left behind. In REAS3, the particle velocities are taken individually according to the histograms.\\ 
An unknown influence might arise due to the underlying atmosphere models used by REAS3 and MGMR which are not equal. In REAS, the same atmosphere models are implemented as given in CORSIKA \cite{Heck1998}. For the present prototype showers, the US standard atmosphere was used. The atmosphere model implemented in MGMR, follows an exponential function for the whole air shower development:
\begin{equation} \label{eq:atmosphereMGMR}
X\left[h\right] = \frac{1000.0}{\cos\theta}\unit[]{\frac{g}{cm^2}} \cdot \exp\left[\frac{\log(0.68)}{4000.0}\frac{h}{\unit[]{m}}\right],
\end{equation}
where $\theta$ is the zenith angle and $h$ is the height above sea level from where the signal is emitted. To calculate the radio emission in REAS3 with a modified atmosphere, the atmosphere model would have to be adapted in CORSIKA as well, since this influences the air shower development. Thus, the modified atmosphere has not been considered for this comparison.\\
Note, that for the comparison in this article, the refractive index of the atmosphere was set to unity for both models since we want to achieve a fundamental understanding of the differences and similarities of MGMR and REAS3. Considering the refractive index, the complexity of the radio emission mechanism would highly increase.

\section{Direct Comparison}\label{sec:directcomp}

In table \ref{tab:prototypeShowers}, the set of selected prototype air showers are listed where the angles correspond to the coordinate system given by CORSIKA, i.e. an azimuth angle of $0^\circ$ corresponds to an air shower coming from south and an azimuth of $90^\circ$ corresponds to an air coming from east. The set contains vertical air showers with various primary energies, one inclined air shower and vertical air showers where the magnetic field first was chosen perpendicular to the shower axis and second parallel to the shower axis.
\begin{table}[tb!]
\begin{center}
\begin{minipage}[b]{0.95\textwidth}
{\footnotesize
\caption[Prototype showers selected for the comparison of REAS3 and MGMR]{Overview of the set of prototype air showers which were simulated for a detailed comparison between REAS3 and MGMR. An inclination of $-37^\circ$ corresponds to the magnetic field geometry of the Pierre Auger Observatory. Note, that the azimuth angle is given in the coordinate system of CORSIKA, i.e. the inclined air shower is coming from south-east.\\} \label{tab:prototypeShowers}
\begin{tabular}{>{\centering\arraybackslash}m{0.22\textwidth}>{\centering\arraybackslash}m{0.22\textwidth}>{\centering\arraybackslash}m{0.22\textwidth}>{\centering\arraybackslash}m{0.22\textwidth}}\hline
\textbf{Energy} & \textbf{Zenith, Azimuth} & \textbf{Strength of $\vec{B}$} & \textbf{Inclination of $\vec{B}$} \\ \hline
$10^{17}$\,eV & $0^\circ$, $0^\circ$   & 0.23\,Gauss & $-37^\circ$ \\
$10^{18}$\,eV & $0^\circ$, $0^\circ$   & 0.23\,Gauss & $-37^\circ$ \\
$10^{19}$\,eV & $0^\circ$, $0^\circ$   & 0.23\,Gauss & $-37^\circ$ \\
$10^{17}$\,eV &	$0^\circ$, $0^\circ$   & 0.23\,Gauss & $0^\circ$ (horizontal)\\
$10^{17}$\,eV & $0^\circ$, $0^\circ$   & 0.23\,Gauss & $90^\circ$ (vertical) \\
$10^{17}$\,eV & $0^\circ$, $0^\circ$   & 0.0\,Gauss  &  - \\
$10^{17}$\,eV & $50^\circ$, $45^\circ$ & 0.23\,Gauss & $-37^\circ$  \\ \hline
\end{tabular}
}
\end{minipage}
\end{center}
\end{table}
One shower was calculated in the absence of any magnetic field to compare the emission of non-geomagnetic radiation. This section focuses on three different points: vertical air showers (section \ref{subs:vertical}), an inclined air shower (section \ref{subs:inclined}) and special magnetic field configurations (section \ref{subs:magnetic}).

\subsection{Emission from vertical air showers}\label{subs:vertical}

To compare the emission from a vertical air shower, three different primary energies were simulated. The primary energies were chosen as $\unit[10^{17}]{eV}$, $\unit[10^{18}]{eV}$ and $\unit[10^{19}]{eV}$ since these are the typical energies measured by the radio enhancement of the Pierre Auger Observatory, AERA \cite{FliescherARENA2010}. The higher the energy of the primary particle $E_p$, the more electrons and positrons are generated in an air shower. A rough estimation for the number of charged particles in the shower maximum is given by $N_{\mbox{max}} = E_p/\unit{GeV} $ \cite{Allan1971}. The scaling of the number of particles with the primary particle energy is directly reflected in the scaling of the field strength with particle energy for coherent radio emission, i.e. $ N_{\mbox{max}} \propto E_p^{0.96} $ (cf. section 4.4 in \cite{HuegeThesis2004}).
	\begin{figure}[tb!]
	\begin{center}
		\begin{minipage}[b]{0.49\linewidth}
		\centering
		\begin{overpic}[angle = 270, width = 1.0\textwidth]{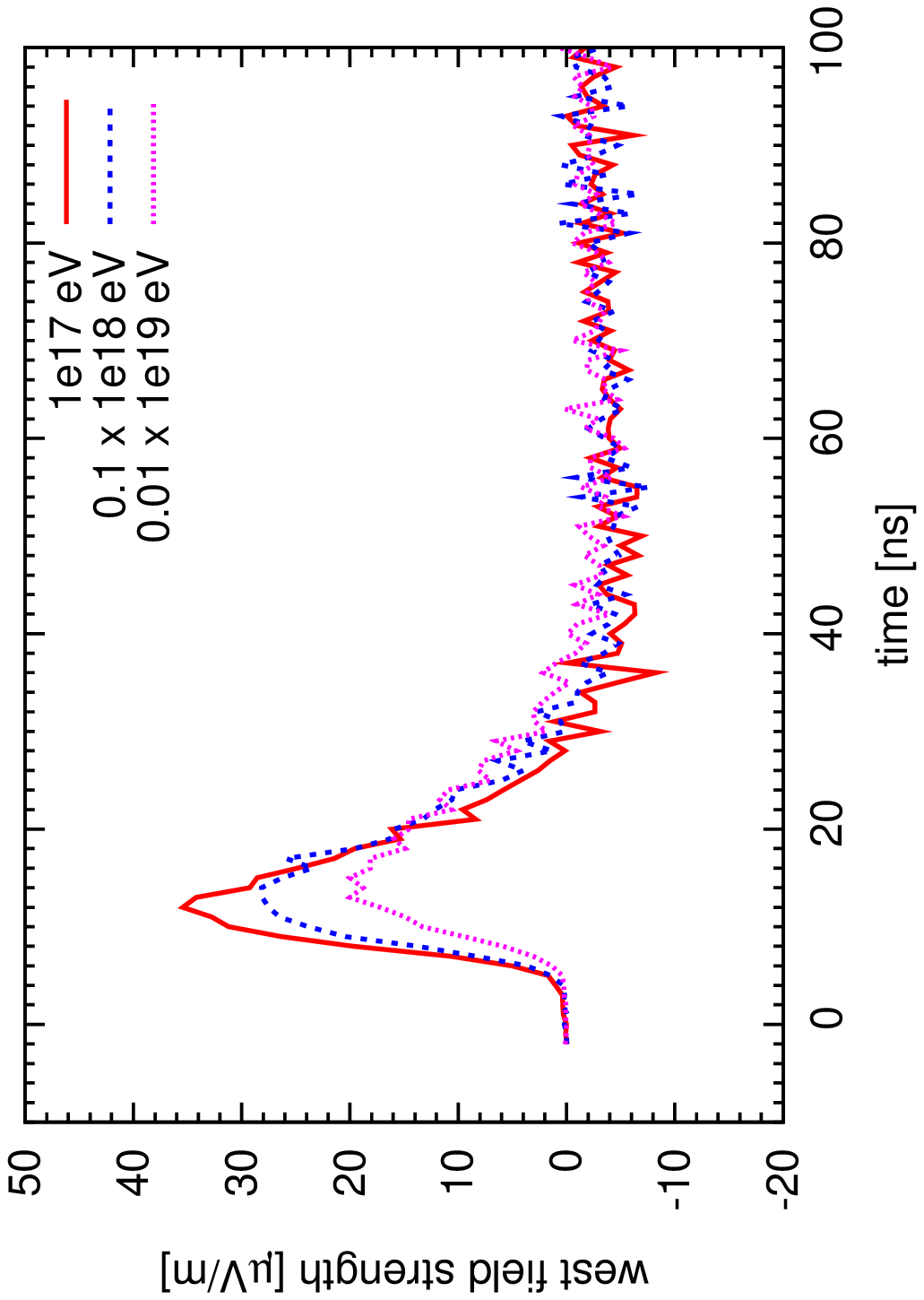}
		\put(25,60){\scriptsize{REAS3}}
		\end{overpic}		
		\end{minipage}
		\begin{minipage}[b]{0.49\linewidth}
		\centering
		\begin{overpic}[angle =270, width = 1.0\textwidth]{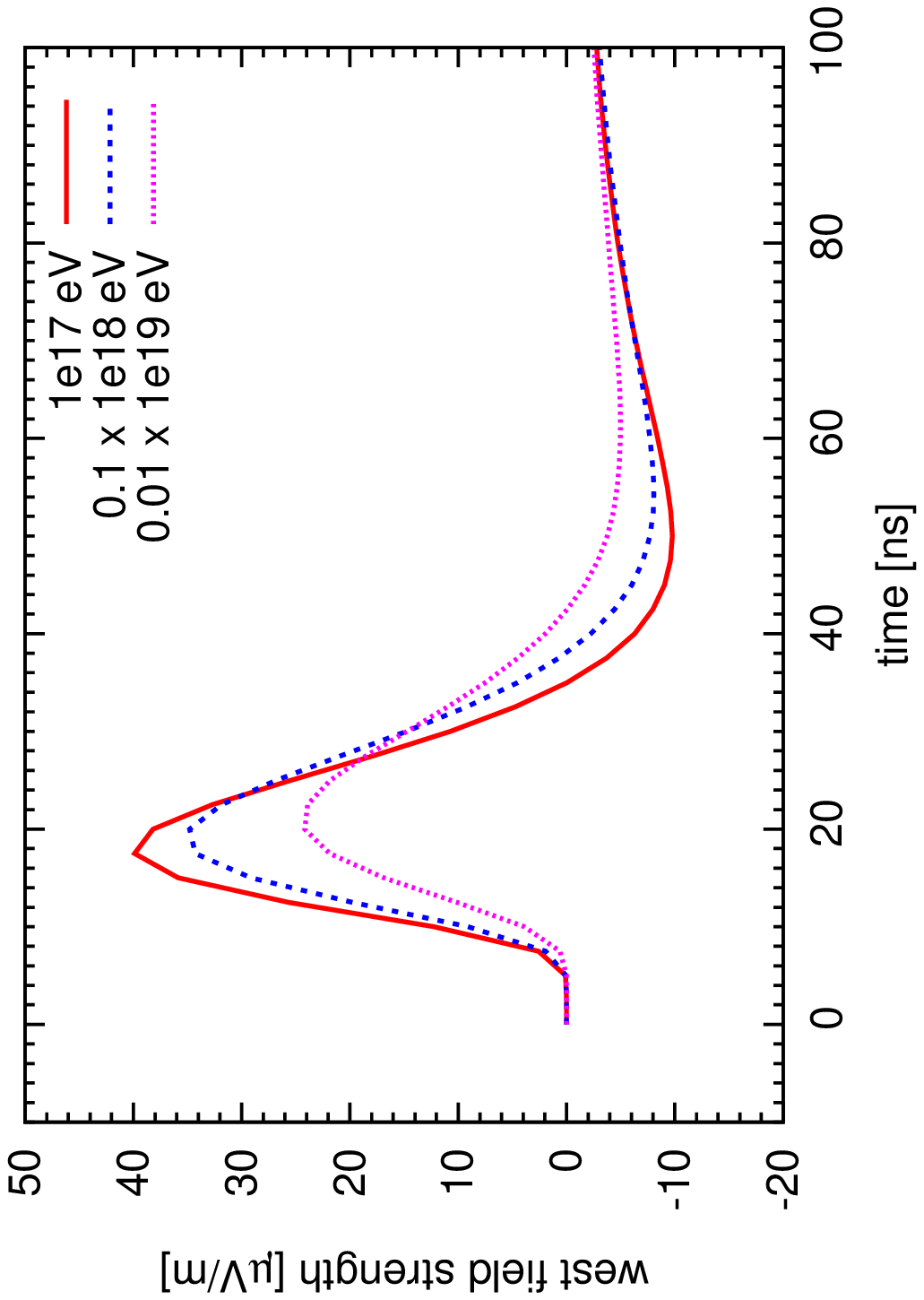}
		\put(25,60){\scriptsize{MGMR}}
		\end{overpic}				
		\end{minipage}
		\caption[Raw pulses for different primary energies (REAS3 and MGMR)]{Comparison of the east-west polarisation component emitted by vertical air showers with three different primary energies at an observer position 200\,m north from the shower core: $\unit[10^{17}]{eV}$ (solid red), $\unit[10^{18}]{eV}$ (dashed blue) and $\unit[10^{19}]{eV}$ (dotted magenta) for REAS3 (left) and MGMR (right). The pulse for $\unit[10^{18}]{eV}$ is multiplied with 0.1 and for $\unit[10^{19}]{eV}$ with 0.01 to allow a better comparison within the same plot. REAS3 and MGMR obtain similar results.}\label{fig:pulsesEnergy}
		\end{center}
	\end{figure}
This increase of the amplitudes with larger primary energy is true for the MGMR model as well as for the REAS3 simulation as shown in figure \ref{fig:pulsesEnergy}. It is obvious that the height of the radio signal approximately (but not exactly) scales linearly with the energy. This is well understood: 
the position of the shower maximum is deeper in the atmosphere for higher primary particle energies and thus, the lateral distribution of the radio signal gets steeper \cite{HuegeFalcke2005}. Combining the dependence of the primary particle energy and the position of the shower maximum, the dependence of the field strength on the primary particle energy is still describable by a power-law. 

In fig. \ref{fig:pulsesEnergy}, the pulses of the vertical air showers with different primary energies are shown. Since the characteristics of the radio signal are mostly unchanged with higher energies \cite{Lafebre2009} and thus the result of the comparison between both models are not influenced by the choice of energy, the following comparison is concentrated on the vertical air shower with primary energy $\unit[10^{17}]{eV}$.

Figure \ref{fig:pulsesEnergy} also shows that the numerical noise level of REAS3 is somewhat higher than the noise level of the MGMR simulation. This effect is mostly relevant for near-vertical showers as discussed in this section. The figures of the inclined air shower illustrate this (cf. section \ref{subs:inclined}). In the MGMR model, the motion of particles is averaged at the beginning of the calculation of the coherent emission and the corresponding electric field is calculated at the end. Thus, the result of the MGMR model is less affected by numerical noise.

\begin{figure}[tb!]
	\begin{center}
		\begin{minipage}[b]{0.49\linewidth}
		\centering
		\begin{overpic}[angle = 270, width = 1.0\textwidth]{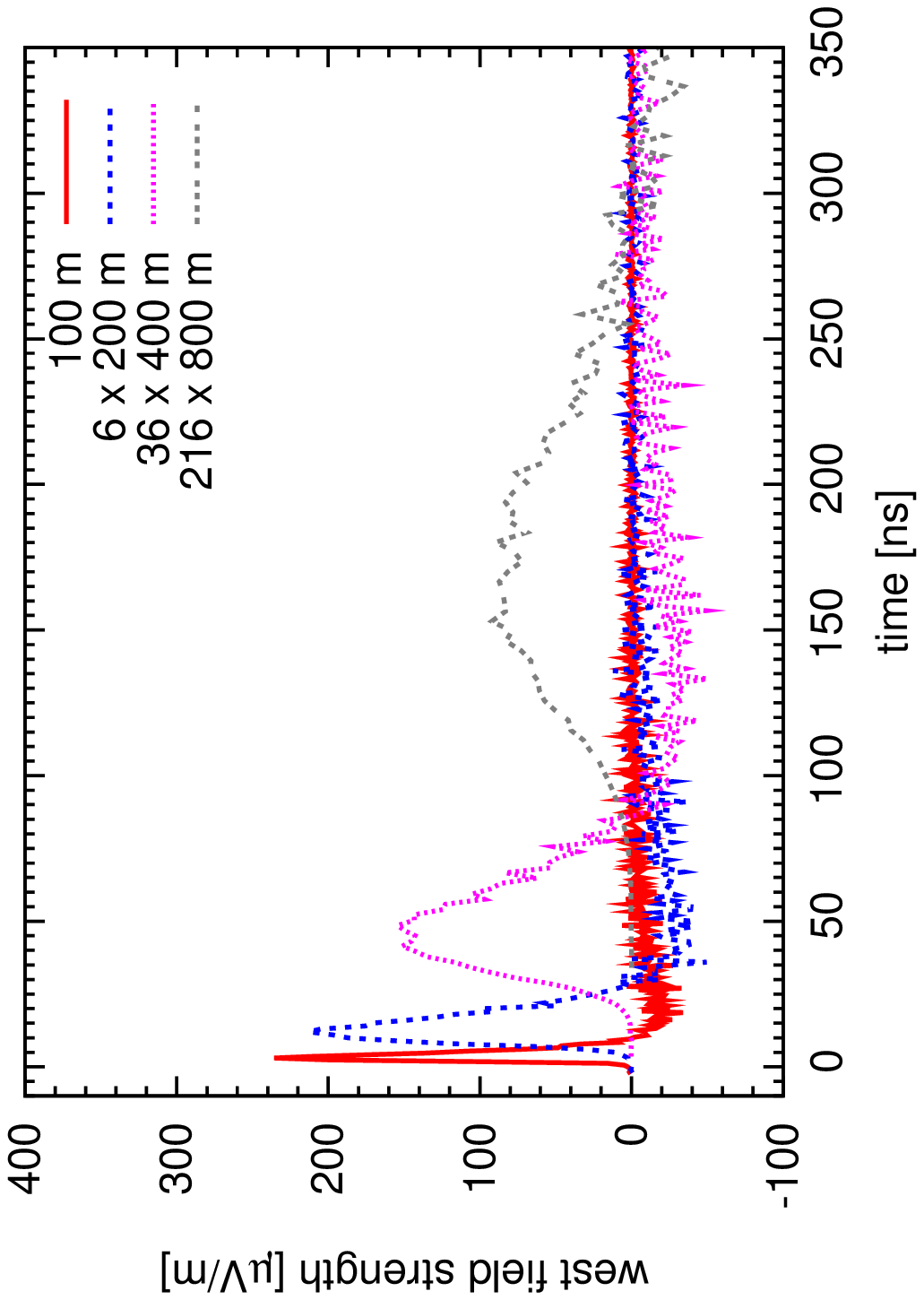}
		\put(25,60){\scriptsize{REAS3}}
		\end{overpic}		
		\end{minipage}
		\begin{minipage}[b]{0.49\linewidth}
		\centering
		\begin{overpic}[angle =270, width = 1.0\textwidth]{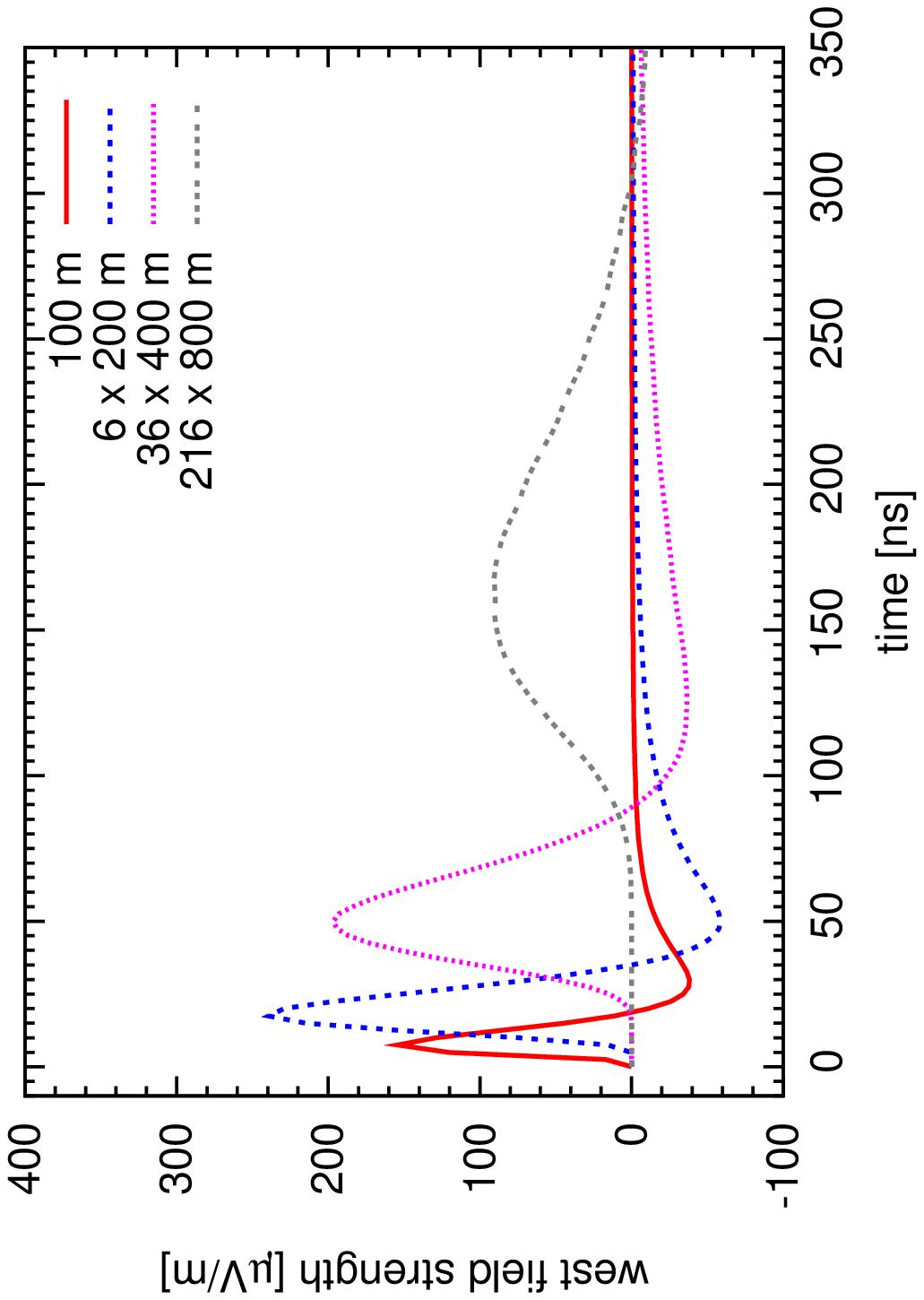}
		\put(25,60){\scriptsize{MGMR}}
		\end{overpic}				
		\end{minipage}
		\caption[Raw pulses at lateral distances (REAS3 and MGMR)]{Comparison of the west polarisation component emitted by a vertical air shower with a primary energy of $\unit[10^{17}]{eV}$ for REAS3 (left) and MGMR (right). The figures show pulses for observers at different lateral distances to the shower core. With increasing distance, the results of both models converge.}\label{fig:pulsesLateral}
		\end{center}
	\end{figure}

Comparing the raw (unlimited bandwidth) radio pulses from the vertical air shower in figure \ref{fig:pulsesLateral}, 
it is obvious that MGMR as well as REAS3 predict bipolar pulses. Furthermore, the amplitudes agree within a factor of $\sim 2$, recalling that the difference between the maximal field strengths of the raw pulses of the previous versions of both models was a factor of ten. Only close to the shower core, the deviations get larger.
     \begin{figure}[tb!]
     \begin{center}
     	\begin{minipage}[b]{0.65\linewidth}
		\centering
		\includegraphics[angle = 270 , width = 1.0\textwidth]{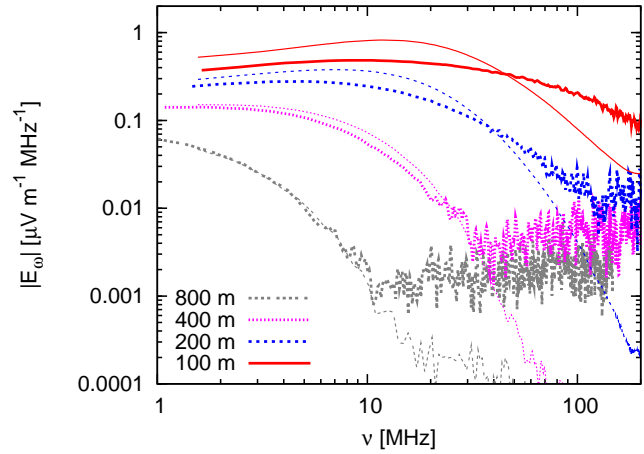}
		\end{minipage}
		\caption[Frequency spectra for REAS3 and MGMR]{Comparison of the frequency spectra for REAS3 (thick lines) and MGMR (thin lines) for a vertical air shower with a primary energy of $\unit[10^{17}]{eV}$. The total spectral field strength is shown for observers at different lateral distances from 100\,m up to 800\,m.} \label{fig:spectraComparison}
		\end{center}
	\end{figure}
This effect can be seen in all other figures presented in this section as well. The remarkable agreement for larger distances is visible in the frequency spectra of figure \ref{fig:spectraComparison} where the total spectral electric field strength is shown as a function of frequency. For observers with lateral distances larger than 400\,m, the results match accurately except for the incoherent noise at high frequencies. At 100\,m observer distance, REAS3 predicts a flatter frequency spectrum than MGMR, especially in the frequency range of $30-80$\,MHz, in which most of the experiments measure.
	\begin{figure}[tb!]
	\begin{center}
		\begin{minipage}[b]{0.49\linewidth}
		\centering
		\begin{overpic}[angle = 270, width = 1.0\textwidth]{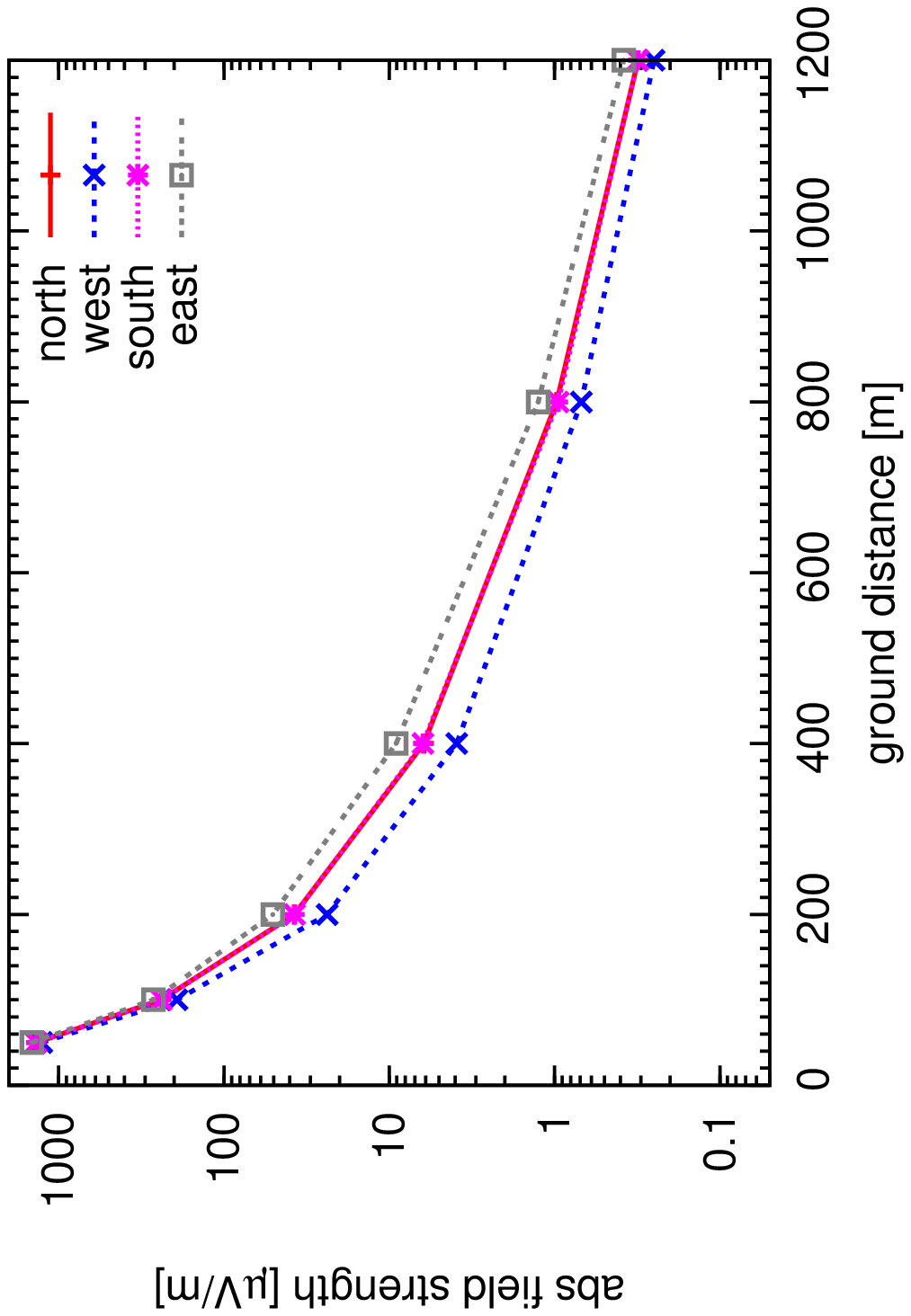}
		\put(25,60){\scriptsize{REAS3}}
		\end{overpic}		
		\end{minipage}
		\begin{minipage}[b]{0.49\linewidth}
		\centering
		\begin{overpic}[angle =270, width = 1.0\textwidth]{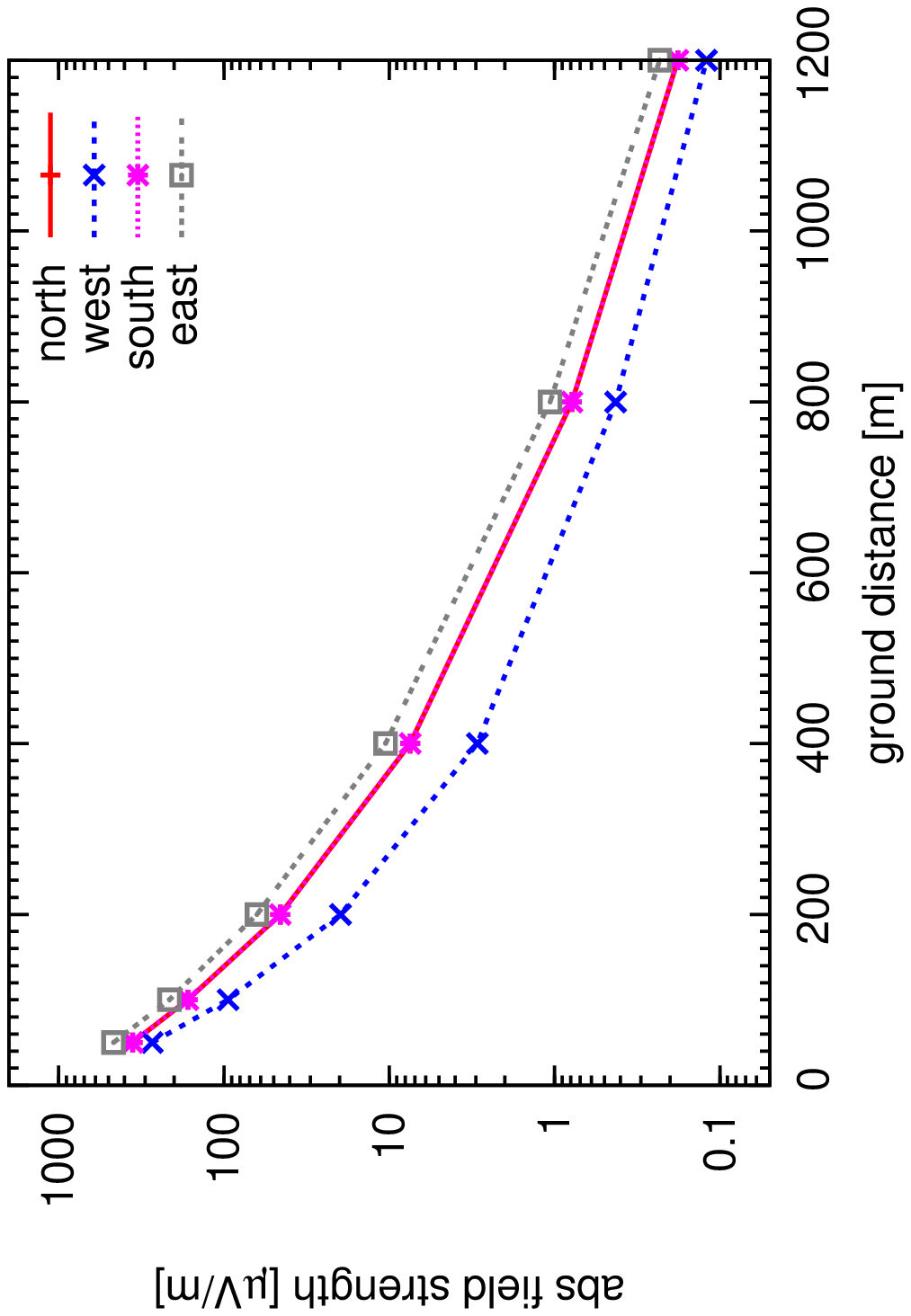}
		\put(25,60){\scriptsize{MGMR}}
		\end{overpic}				
		\end{minipage}
		\caption[Lateral dependence of REAS3 and MGMR]{Comparison of the lateral dependences with full bandwidth amplitudes for a vertical air shower with a primary energy of $\unit[10^{17}]{eV}$ predicted by REAS3 (left) and MGMR (right). The figures display the maximum absolute field strength at a given lateral distance of an observer to the shower axis.}\label{fig:lateralAbsolute}
		\end{center}
	\end{figure}
The predicted amplitude of the radio emission in different azimuthal directions of the air shower can be compared with the lateral distributions of the peak amplitude of unlimited bandwidth pulses, i.e. the unfiltered signal, as illustrated in figure \ref{fig:lateralAbsolute}. The lateral distributions derived from REAS3 and MGMR show an evident east-west-asymmetry. This asymmetry is explained by the time-varying charge excess in air showers causing a radiation contribution with a radial polarisation signature (cf. \cite{Ludwig2010,deVries2010}).
Therefore, the two models give evidence that the radio emission from cosmic ray air showers is not purely $ \vec{v}\times\vec{B} $ polarised.
 	
\begin{figure}[tb!]
	\begin{center}
		\hspace{-0.8cm}
		\begin{minipage}[b]{0.34\linewidth}
		\centering
		\includegraphics[angle = 270 , width = 1.2\textwidth]{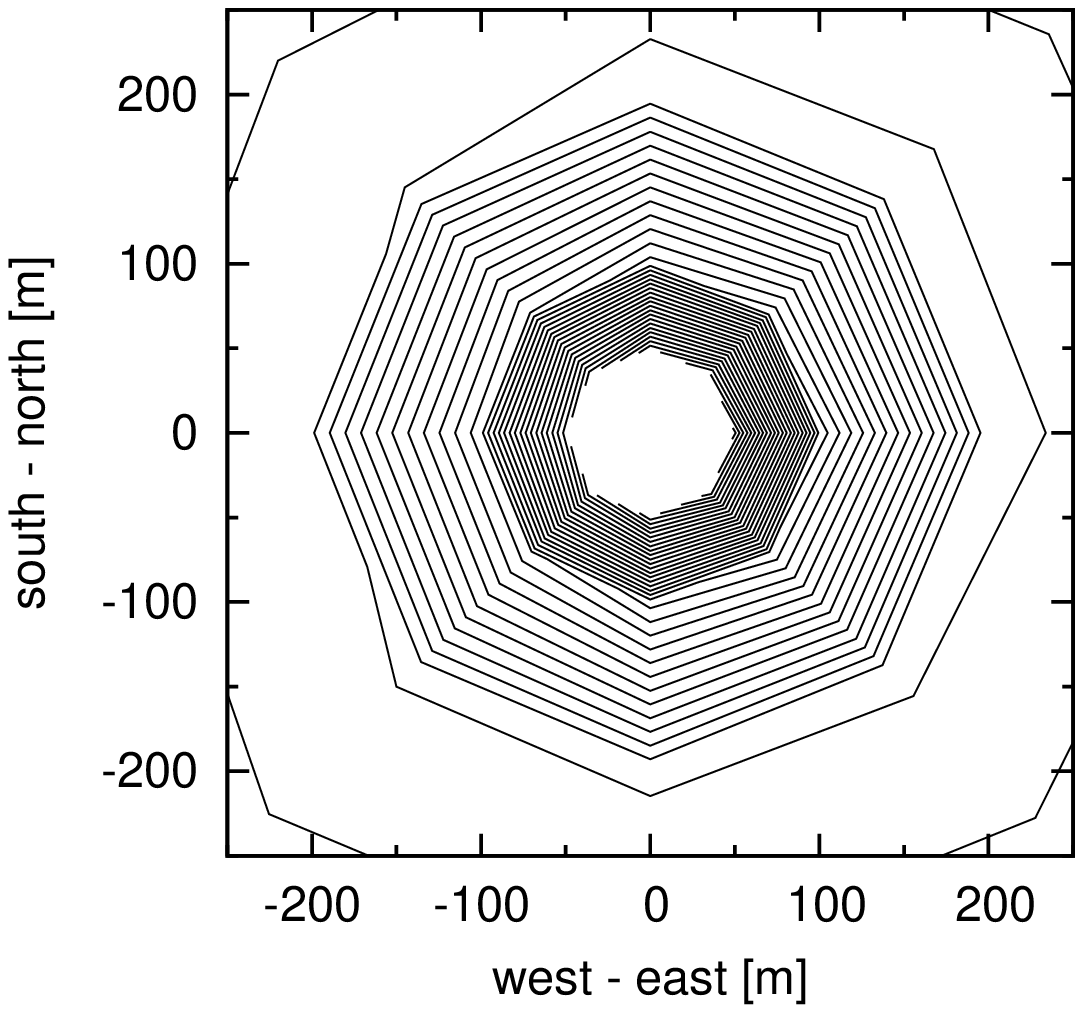}
		\includegraphics[angle = 270 , width = 1.2\textwidth]{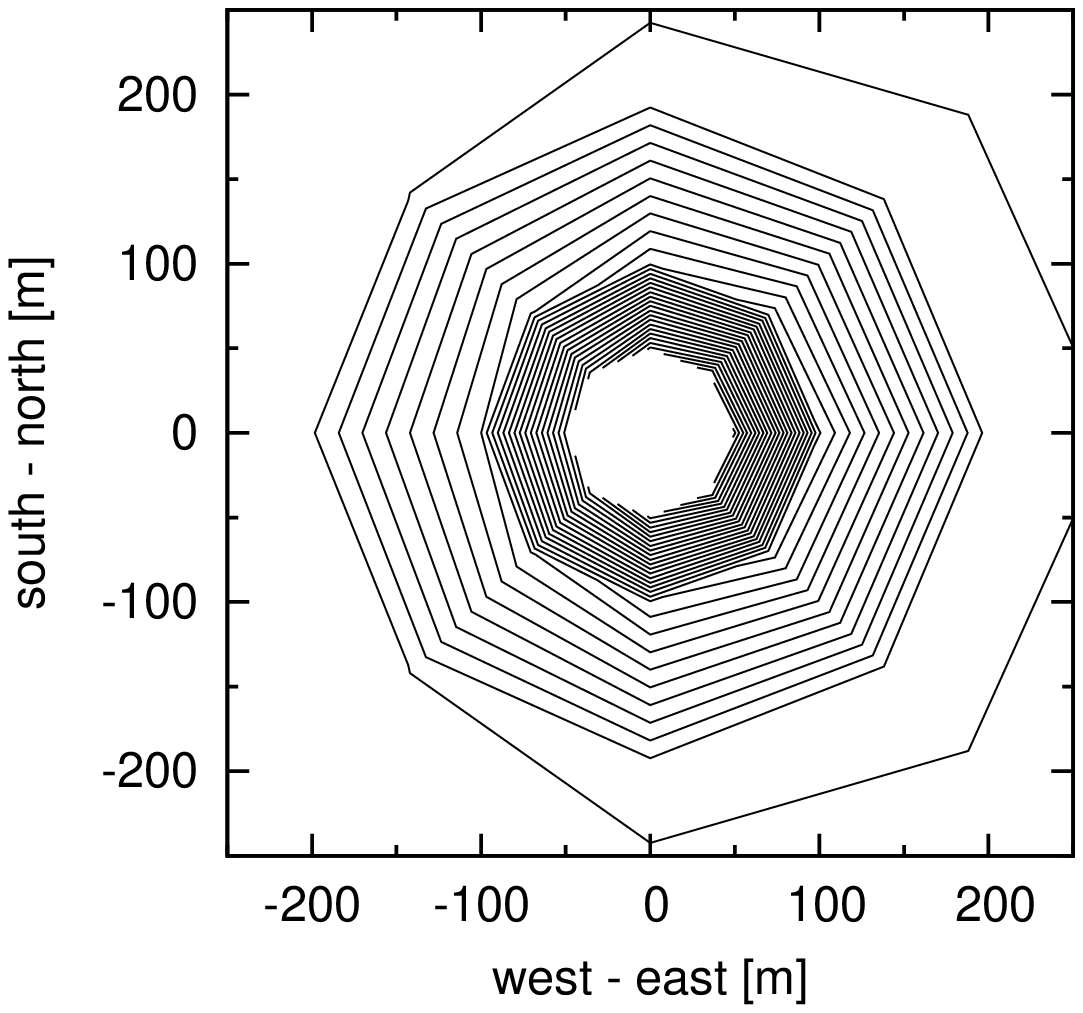}
		\end{minipage}
		\hspace{-0.6cm}
		\begin{minipage}[b]{0.34\linewidth}
		\centering
		\includegraphics[angle = 270 , width = 1.2\textwidth]{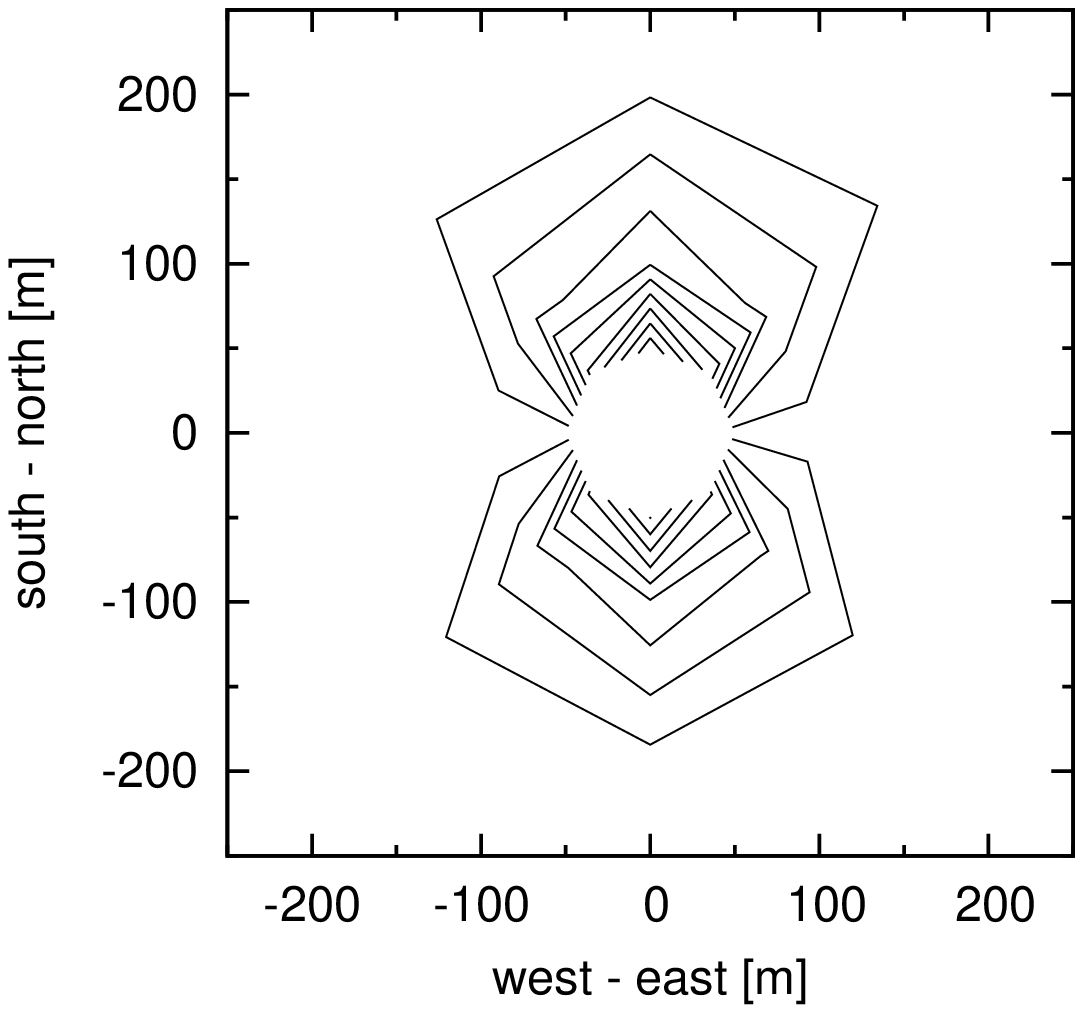}
		\includegraphics[angle = 270 , width = 1.2\textwidth]{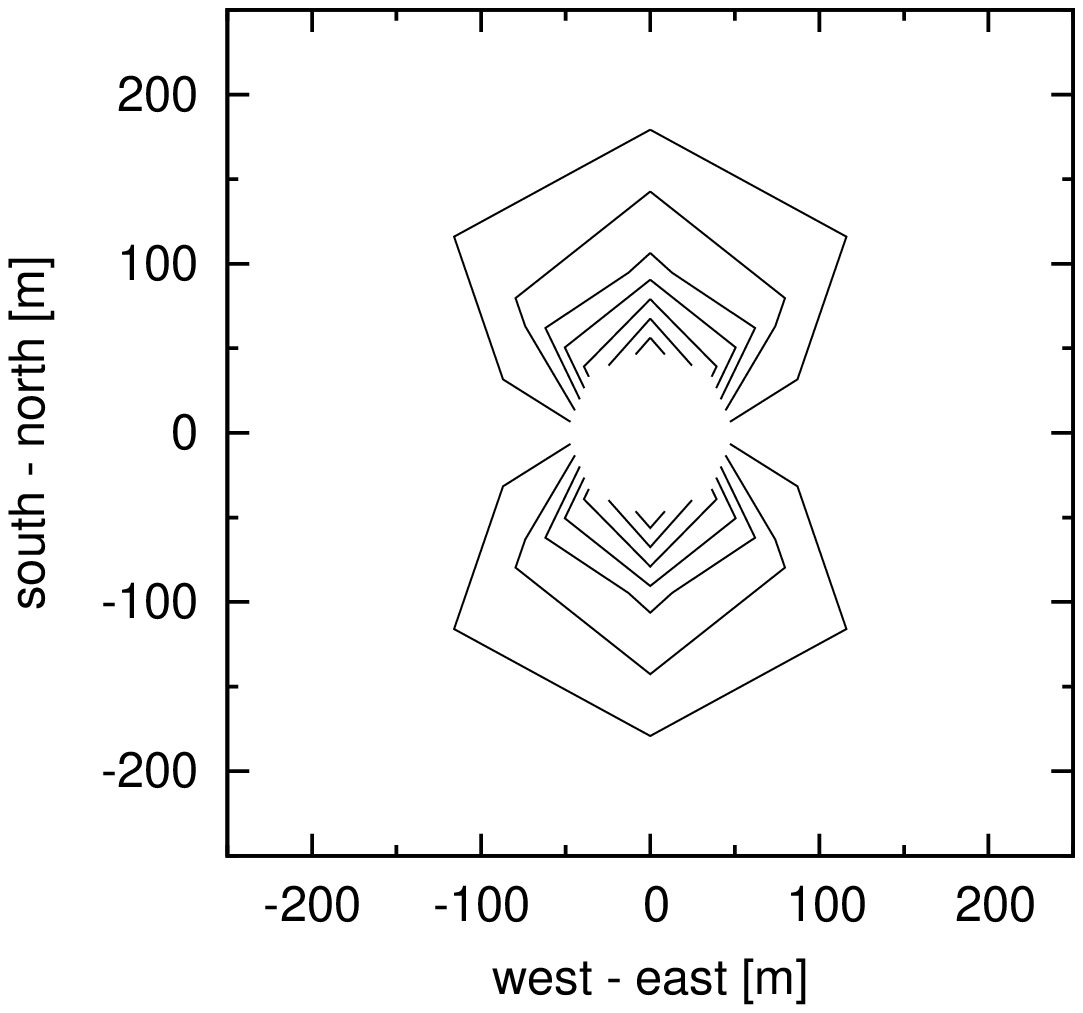}
		\end{minipage}
		\hspace{-0.6cm}
		\begin{minipage}[b]{0.34\linewidth}
		\centering
		\includegraphics[angle = 270 , width = 1.2\textwidth]{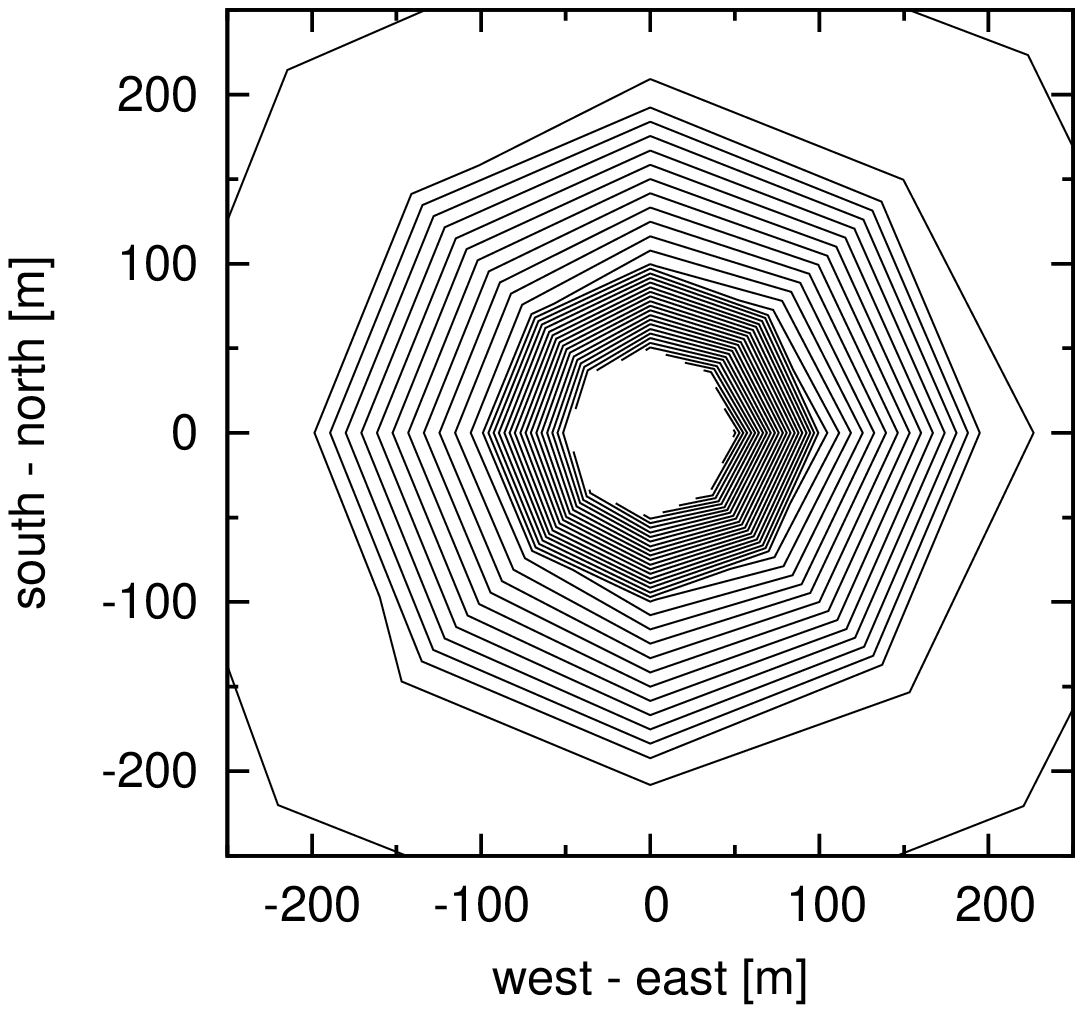}
		\includegraphics[angle = 270 , width = 1.2\textwidth]{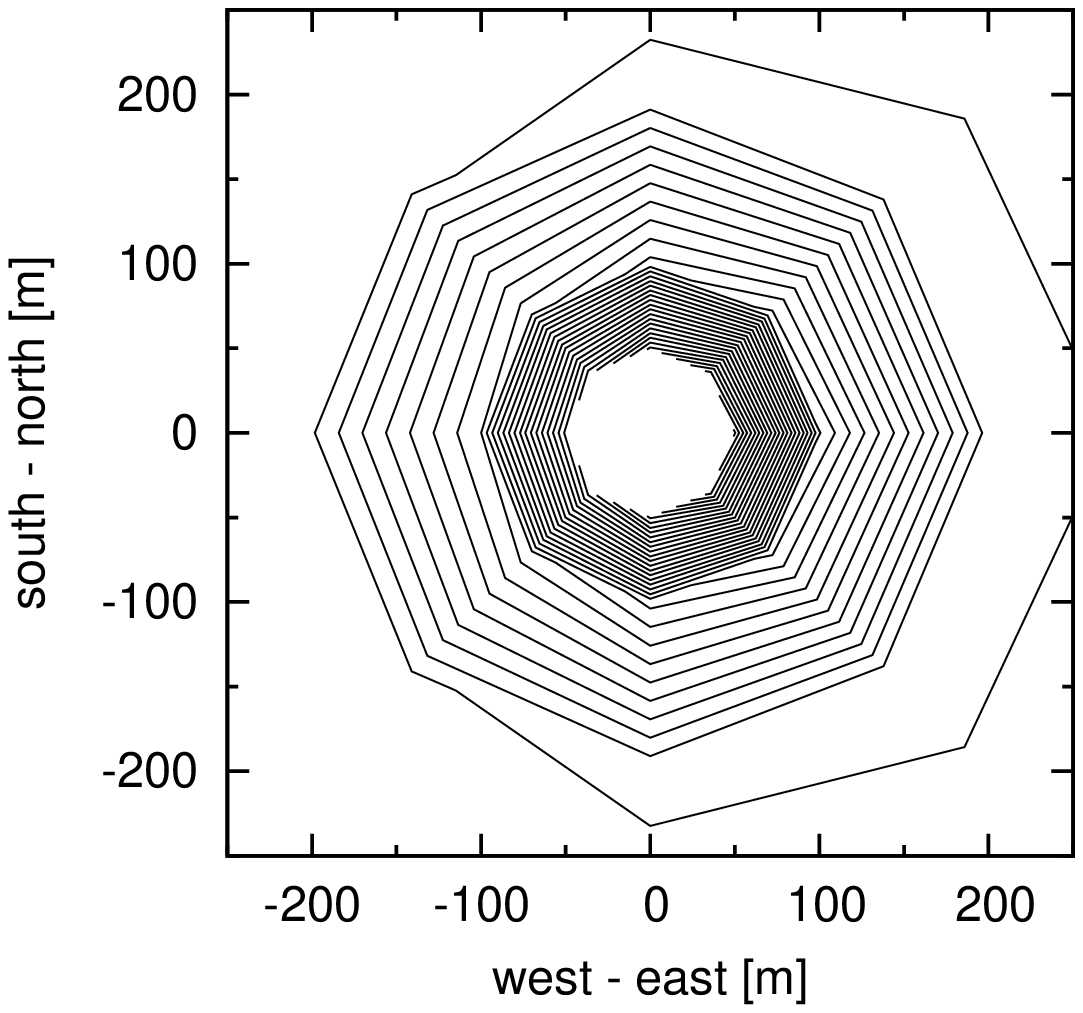}
		\end{minipage}
		\caption[Contour plots of REAS3 and MGMR (vertical shower)]{Contour plots of the 60\,MHz field strength for the emission from a $10^{17}$\,eV vertical air shower.
		From left to right:	total field strength, north-south and east-west polarisation component. 
		Contour levels are 0.1\,$\mu$Vm$^{-1}$MHz$^{-1}$ apart. The closest position of the simulated observers to the shower core
		is 50\,m. Upper row: REAS3. Lower row: MGMR} \label{fig:contourComparison}
		\end{center}
	\end{figure}

Looking at the contour plots of the 60\,MHz spectral emission component displayed in figure \ref{fig:contourComparison}, the same signature is visible. For radiation due to pure geomagnetic emission, no contribution for the north-south polarisation would be expected for the vertical air shower, as well as a symmetric pattern in the east-west polarisation component. The comparison of the contour plots, however, shows that there are still deviations between MGMR and REAS3, in particular in the strength of the east-west asymmetry predicted by the models. 

\subsection{Emission from an inclined air shower}\label{subs:inclined}

\begin{figure}[tb!]
	\begin{center}
		\begin{minipage}[b]{0.49\linewidth}
		\centering
		\begin{overpic}[angle = 270, width = 1.0\textwidth]{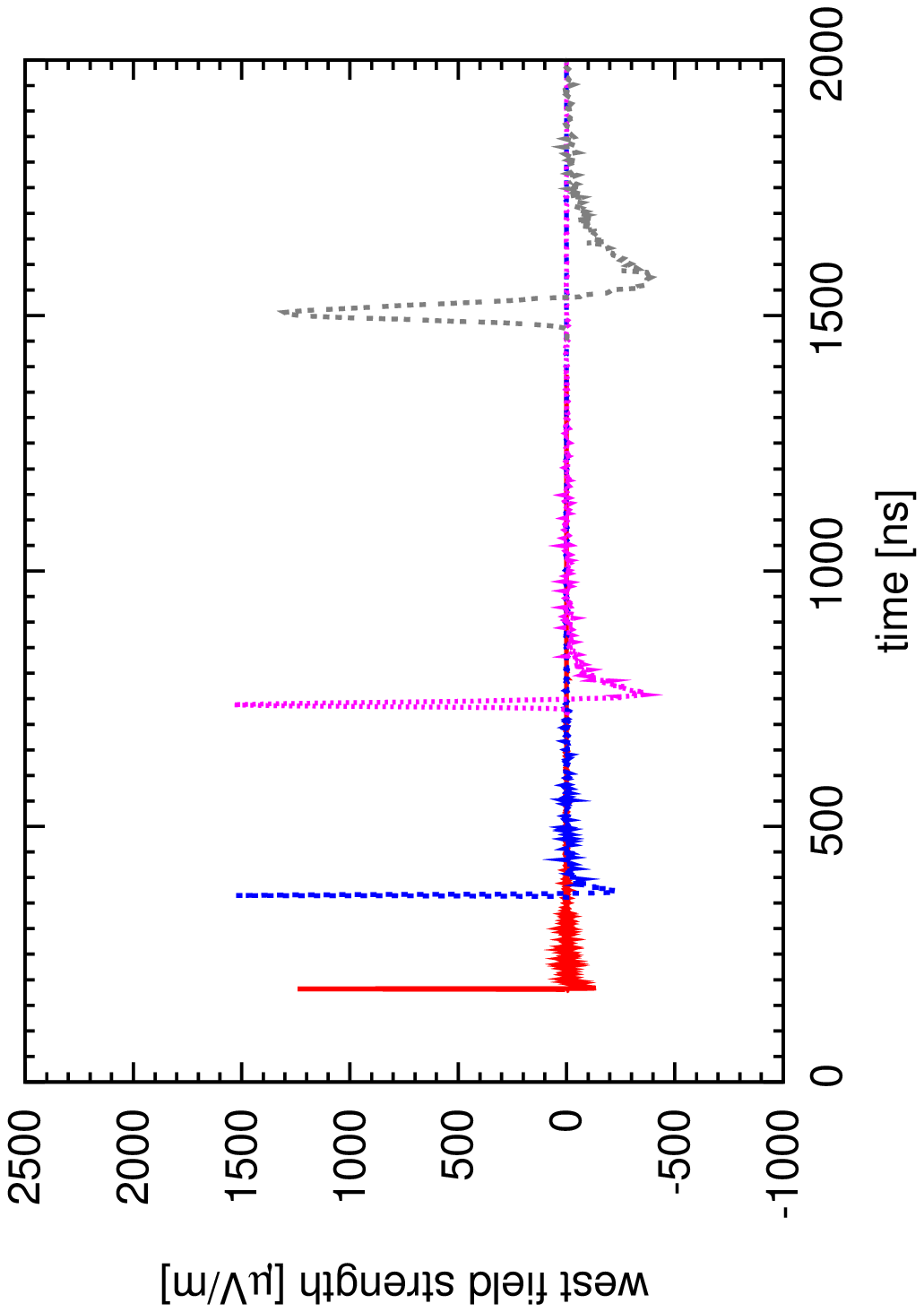}
		\put(30,60){\scriptsize{REAS3}}
		\end{overpic}		
		\end{minipage}
		\begin{minipage}[b]{0.49\linewidth}
		\centering
		\begin{overpic}[angle =270, width = 1.0\textwidth]{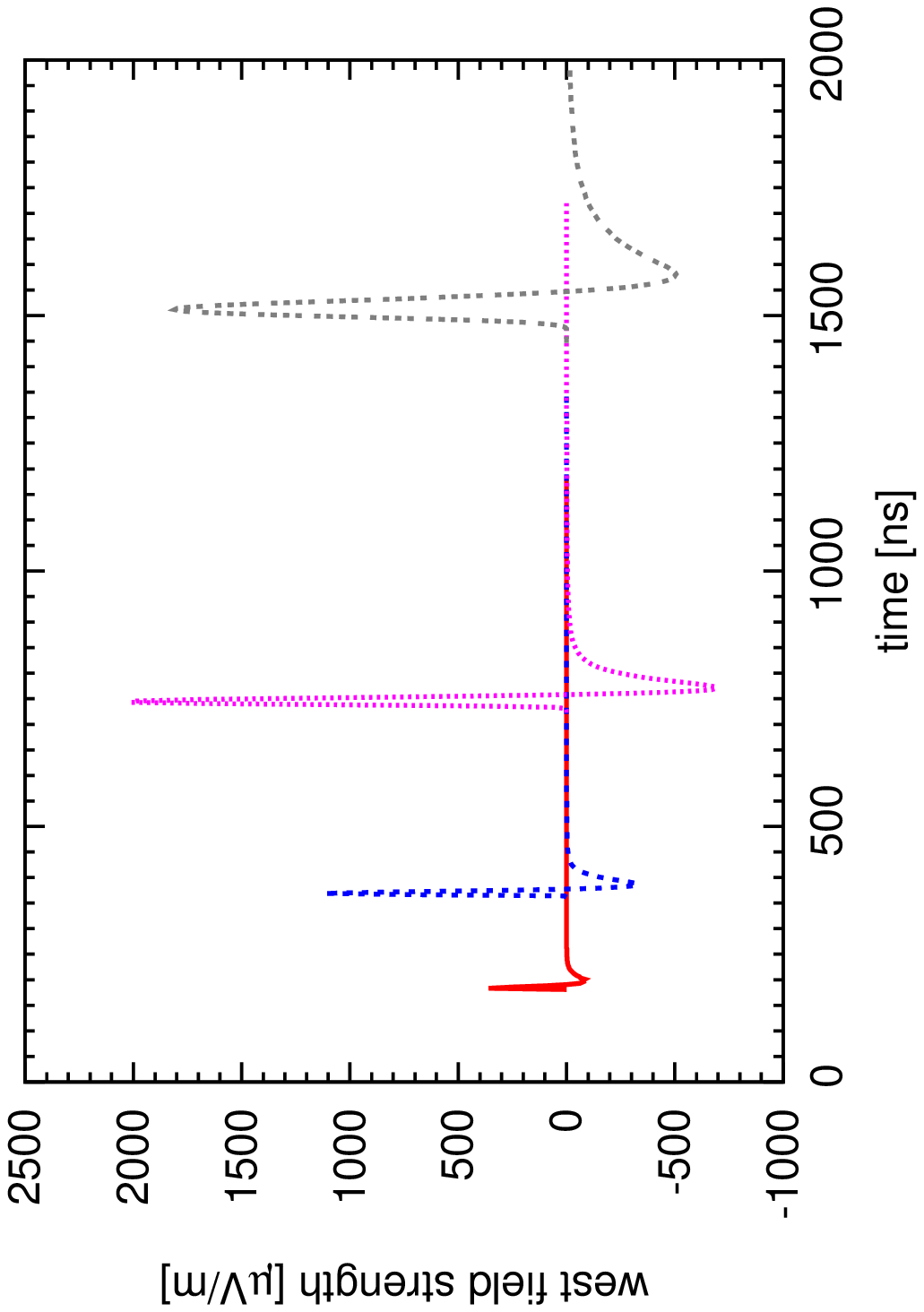}
	   	\put(30,60){\scriptsize{MGMR}}
		\end{overpic}				
		\end{minipage}
		\caption[Raw pulses at lateral distances (inclined shower)]{Comparison of the east-west polarisation component emitted by a 50$^\circ$ inclined air shower with a primary energy of $\unit[10^{17}]{eV}$ for REAS3 (left) and MGMR (right). The figures show pulses for observers at different lateral distances to the shower core. With increasing distance, the results converge. For small distances, the predictions of both models differ by a factor of three.}\label{fig:pulsesLateral50deg}
		\end{center}
	\end{figure}

\begin{figure}[tb!]
	\begin{center}
		\begin{minipage}[b]{0.49\linewidth}
		\centering
		\begin{overpic}[angle = 270, width = 1.0\textwidth]{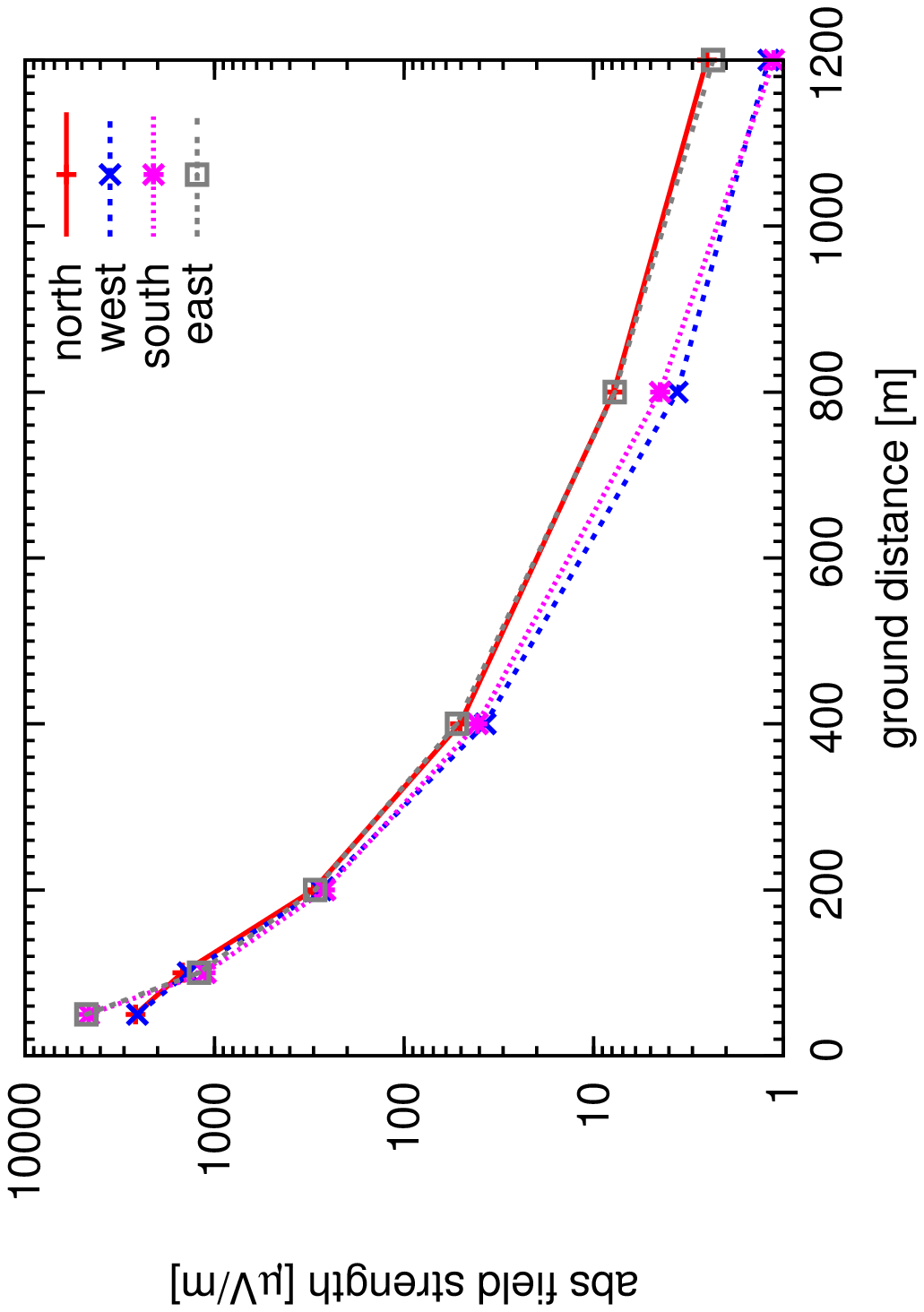}
		\put(30,60){\scriptsize{REAS3}}
		\put(30,55){\scriptsize{N$\approx$E, S$\approx$W}}
		\end{overpic}		
		\end{minipage}
		\begin{minipage}[b]{0.49\linewidth}
		\centering
		\begin{overpic}[angle =270, width = 1.0\textwidth]{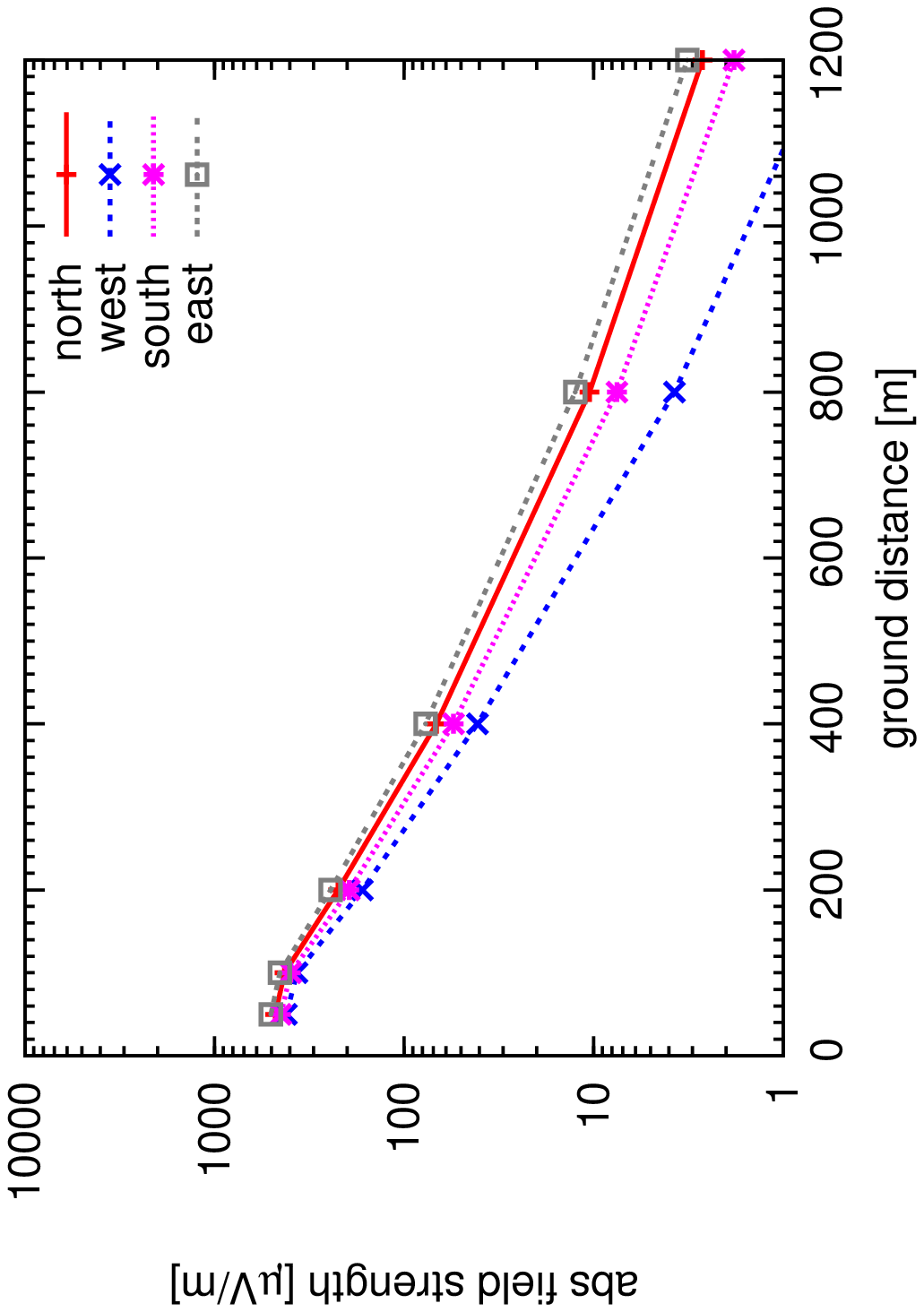}
		\put(30,60){\scriptsize{MGMR}}
		\put(30,55){\scriptsize{N$\neq$E, S$\neq$W}}		
		\end{overpic}				
		\end{minipage}
		\caption[Lateral dependence of REAS3 and MGMR (inclined shower)]{Comparison of the lateral dependences with full bandwidth amplitudes for an air shower with zenith angle of 50$^\circ$ and primary energy of $\unit[10^{17}]{eV}$ predicted by REAS3 (left) and MGMR (right). The figures display the maximum absolute field strength at a given lateral distance to the shower core for observers along various azimuthal directions.}\label{fig:lateralAbsolute50deg}
		\end{center}
	\end{figure}

In this section, the results for an inclined air shower with primary energy of $10^{17}$\,eV and a zenith angle of 50$^\circ$ are compared. The azimuth angle of 45$^\circ$ denotes that the shower is coming from south-east (i.e., pointing to north-west). Figure \ref{fig:pulsesLateral50deg} shows the raw pulses simulated with both models for this geometry. Please note that the zero time corresponds to the time when the primary particle would hit the ground. Since the observers are located north of the shower core and the air shower is coming from south-east, the emission arrives later than in the vertical case.
Close to the shower core, the predictions of REAS3 and MGMR differ nearly by a factor of three, whereas the results are almost the same for larger lateral distances. The pulses derived by REAS3 close to the shower core exhibit higher amplitudes than the pulses obtained with MGMR. The larger deviations close to the shower core are also evident in figure \ref{fig:lateralAbsolute50deg}, where the lateral distributions are shown. 	
The amplitudes predicted by REAS3 increase significantly with smaller distances to the shower axis. For MGMR, the lateral distribution is somewhat flattening to the center. Furthermore, the asymmetries between each azimuthal observer direction, i.e. north, east, south and west, are larger for MGMR than for REAS3. 
	\begin{figure}[htb]
	\begin{center}
		\begin{minipage}[b]{0.65\linewidth}
		\centering
		\includegraphics[angle = 270, width = 1.0\textwidth]{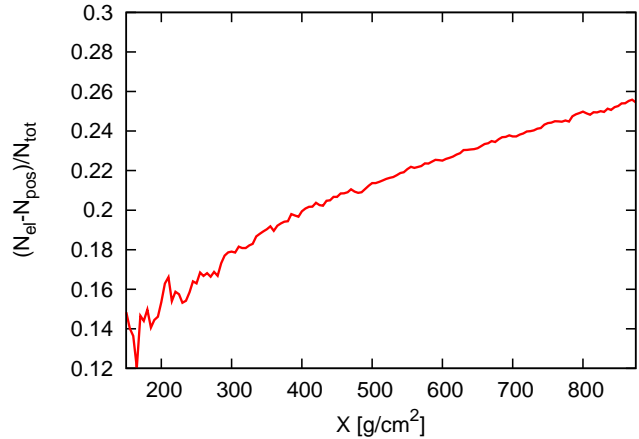}
		\end{minipage}
		\caption[Charge excess profile inclined shower]{Fraction of charge excess with respect to the total number of electrons and positrons as a function of atmospheric depth. With larger atmospheric depth the fraction rises. The large fluctuations below $\unit[200]{\frac{g}{cm^2}}$ are due to small particle statistics.} \label{fig:CEratioInclined}
		\end{center}
	\end{figure}
In REAS3, the observers in the north and east receive nearly the same signal as the observers in the south and west. For MGMR this is not true which might be related to the moving dipole included in MGMR but not in REAS3. The reason for the discrepancies in the azimuthal symmetry might be the different implemenations of the charge excess in MGMR and REAS3. In MGMR, the fraction of the charge excess with respect to the total number of electrons and positrons is assumed to be constant, whereas the longitudinal profile taken into account in REAS3 is changing from roughly $14\%$ to $26\%$ as shown in figure \ref{fig:CEratioInclined}.

Above all, the differing predictions in the amplitudes might be a hint that close to the shower core, the details of the air shower model, which differs in REAS3 and MGMR become important. The air shower model has a larger impact on inclined air showers than on nearly vertical ones, as identical ground distances correspond to smaller effective lateral axis distances.
Moreover, the geometrical distance between observer and shower maximum increases with larger zenith angle. Hence, we discuss the influence of the air shower model in greater depth in section \ref{sec:simplified}.

\subsection{Specific magnetic field configurations}\label{subs:magnetic}

In addition to the realistic air shower geometries shown in the previous sections, it is interesting to look at more contrived situations such as some special magnetic field configurations, since the Earth's magnetic field is responsible for the geomagnetic radio emission in air showers. For this comparison, the magnetic field was once switched off completely and once was chosen to be parallel to the air shower axis. With these special configurations of the magnetic field, the influence of the radiation due to the time-variation of the net charge excess\footnote{This means ``Askaryan radiation'' \cite{Askaryan1962, Askaryan1965} but without Cherenkov-like effects since the refractive index is unity.} is studied.  
	\begin{figure}[tb!]
	\begin{center}
		\begin{minipage}[b]{0.49\linewidth}
		\centering	
		\begin{overpic}[angle = 270, width = 1.0\textwidth]{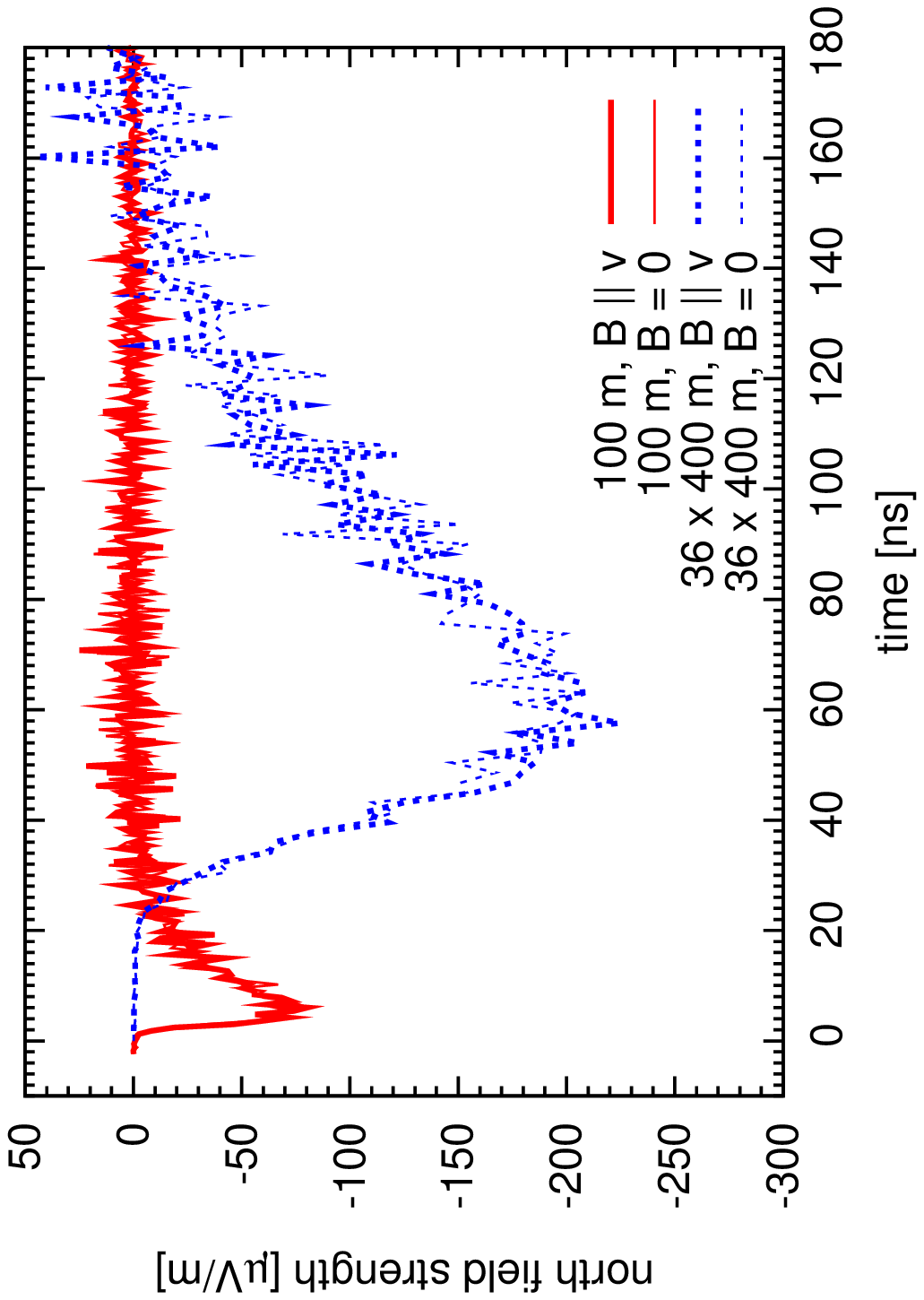}
		\put(25,15){\scriptsize{REAS3}}
		\end{overpic}		
		\end{minipage}
		\begin{minipage}[b]{0.49\linewidth}
		\centering
		\begin{overpic}[angle =270, width = 1.0\textwidth]{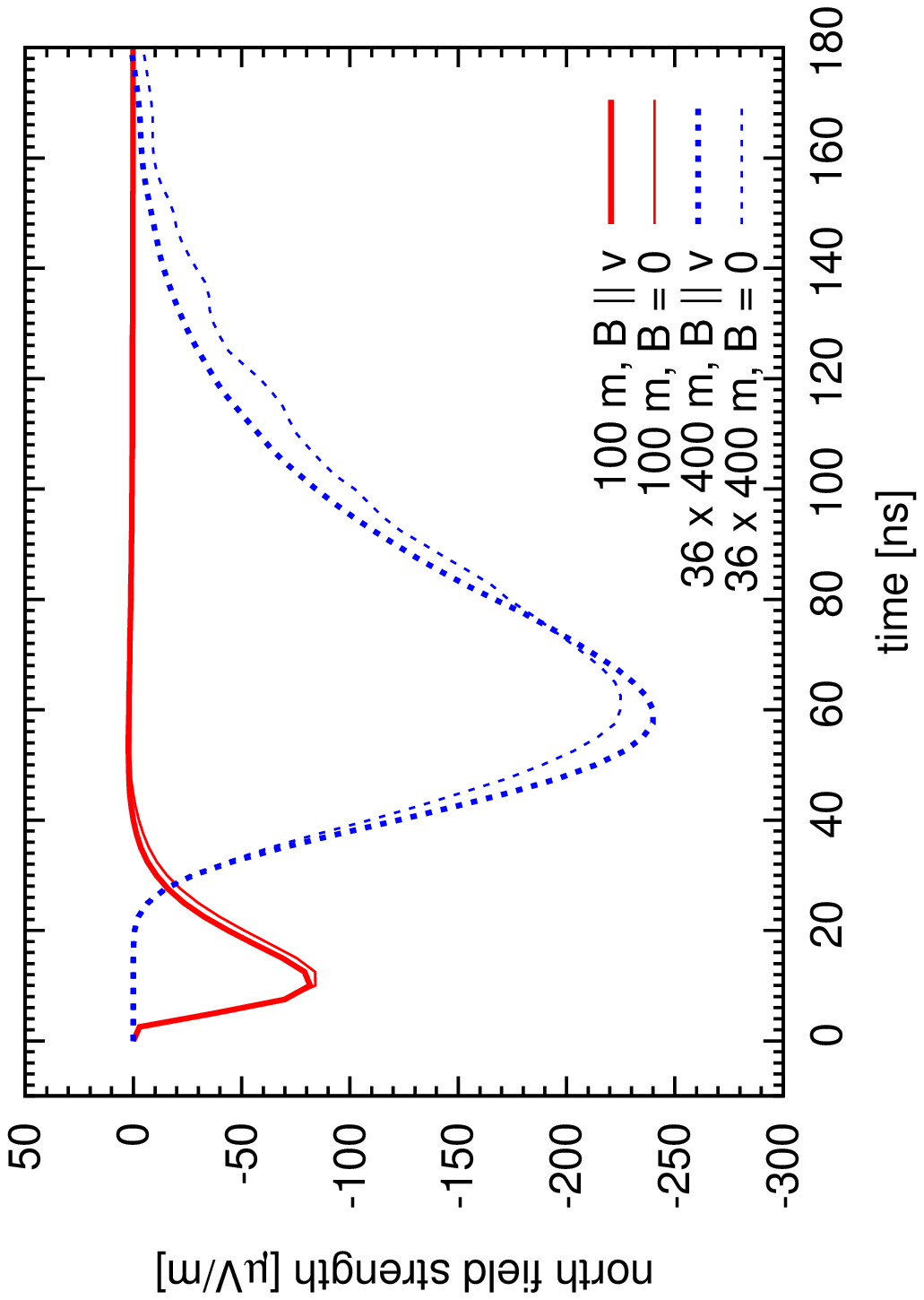}
		\put(25,15){\scriptsize{MGMR}}
		\end{overpic}				
		\end{minipage}
		\caption[Raw pulses for specific B-fields]{Comparison of the pulses emitted by a vertical $ 10^{17}\,$eV air shower in the absence of a magnetic field (thin lines) and a magnetic field parallel to the shower axis (thick lines). Left: REAS3. Right: MGMR.}\label{fig:pulsesNoBfield}
		\end{center}
	\end{figure}
Figure \ref{fig:pulsesNoBfield} shows that the results for these two configurations are indeed similar.
	\begin{figure}[tb!]
	\begin{center}
		\begin{minipage}[b]{0.65\linewidth}
		\centering
		\includegraphics[angle = 270, width = 1.0\textwidth]{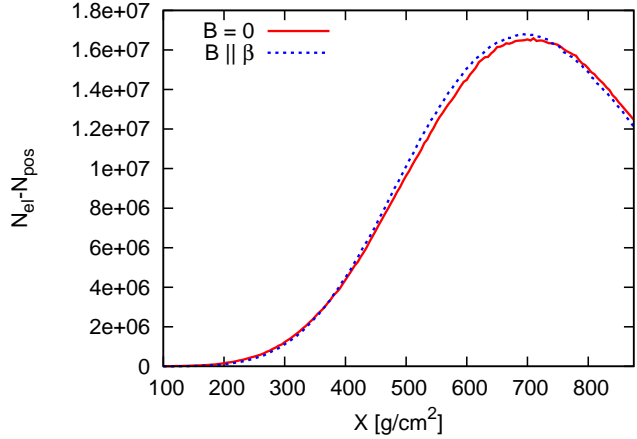}
		\end{minipage}
		\caption[Charge excess profile inclined shower]{Longitudinal development of the charge excess component for the shower without magnetic field (solid red) and for a magnetic field parallel to the shower axis (dashed blue).} \label{fig:CEratioB}
		\end{center}
	\end{figure}
REAS3 and MGMR predict both nearly unipolar pulses. This is not surprising since the longitudinal development of the charge excess component as displayed in figure \ref{fig:CEratioB} reaches the shower maximum only short before the observing height of 1400\,m which corresponds to $\unit[875]{\frac{g}{cm^2}}$ for a vertical air shower. The predictions of REAS3 and MGMR for the pure charge excess case agree very well. The differences in the strength of the pulses are in the order of 10-20\%.
	\begin{figure}[tb!]
	\begin{center}
		\hspace{-0.8cm}
		\begin{minipage}[b]{0.34\linewidth}
		\centering		
		\includegraphics[angle = 270 , width = 1.2\textwidth]{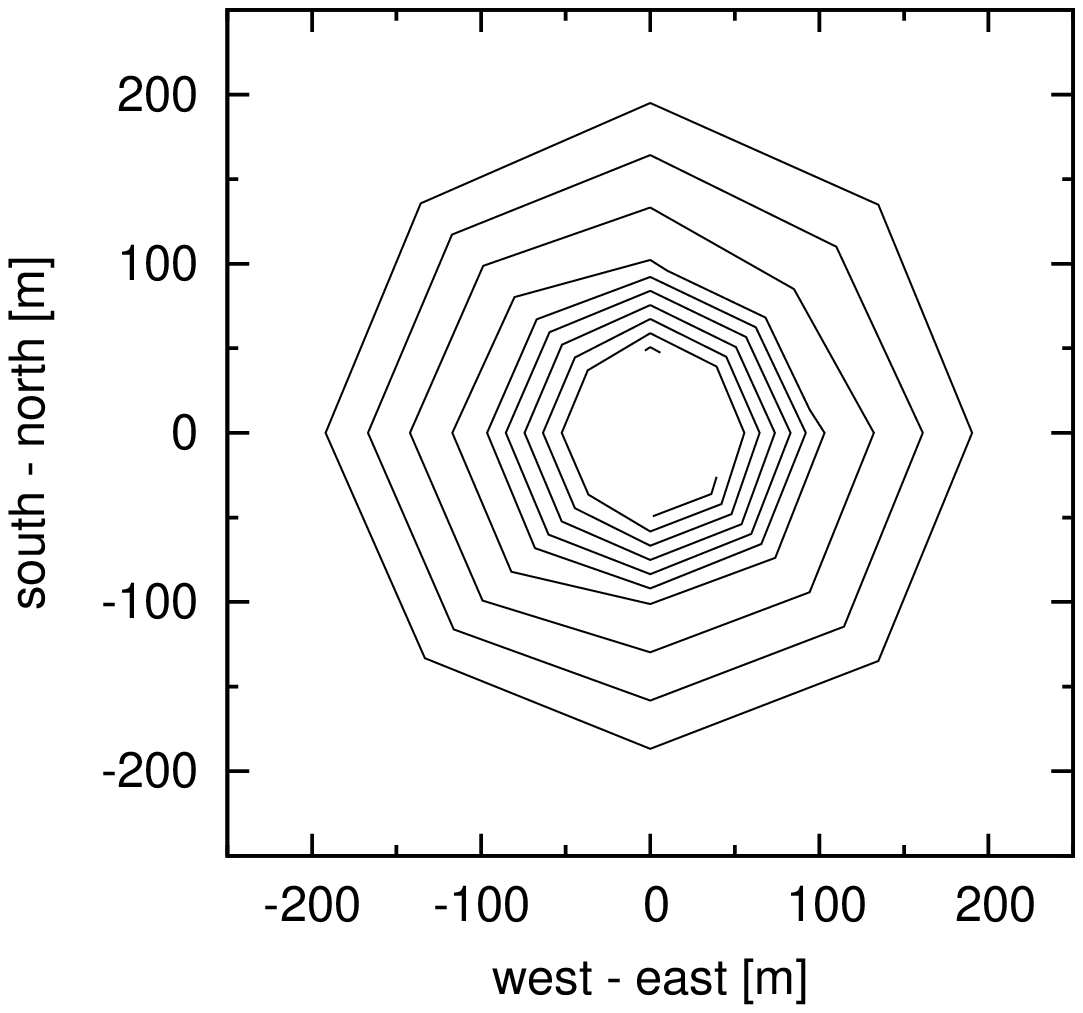}
		\includegraphics[angle = 270 , width = 1.2\textwidth]{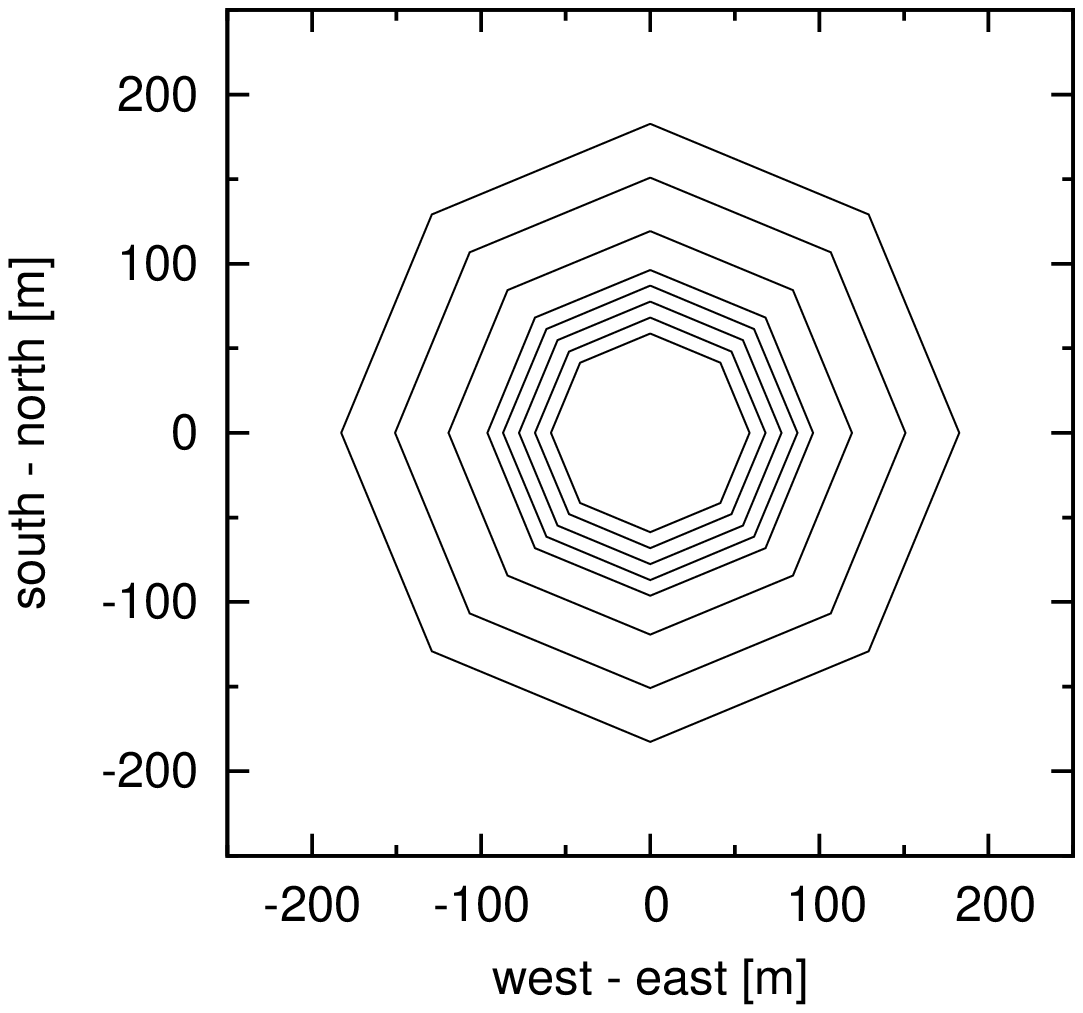}
		\end{minipage}
		\hspace{-0.6cm}
		\begin{minipage}[b]{0.34\linewidth}
		\centering
		\includegraphics[angle = 270 , width = 1.2\textwidth]{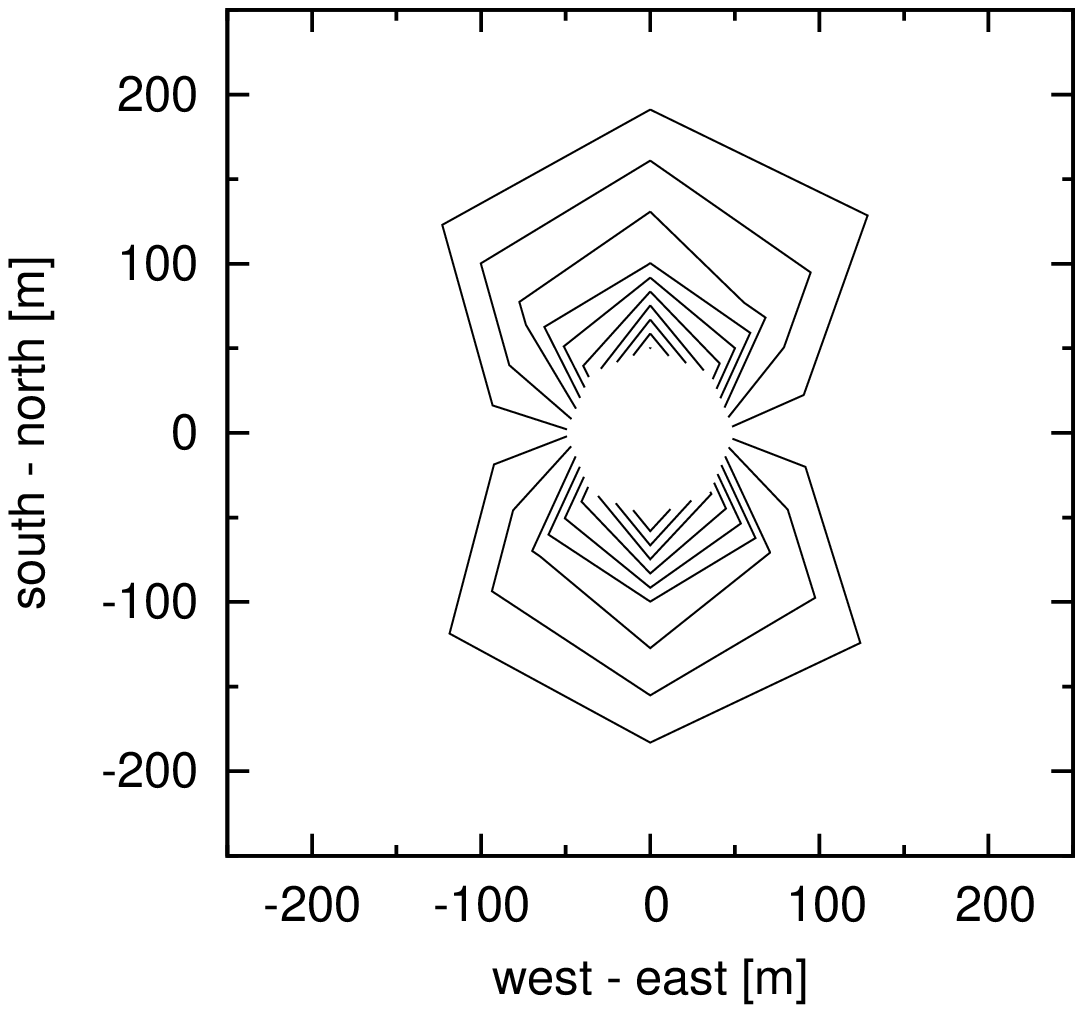}
		\includegraphics[angle = 270 , width = 1.2\textwidth]{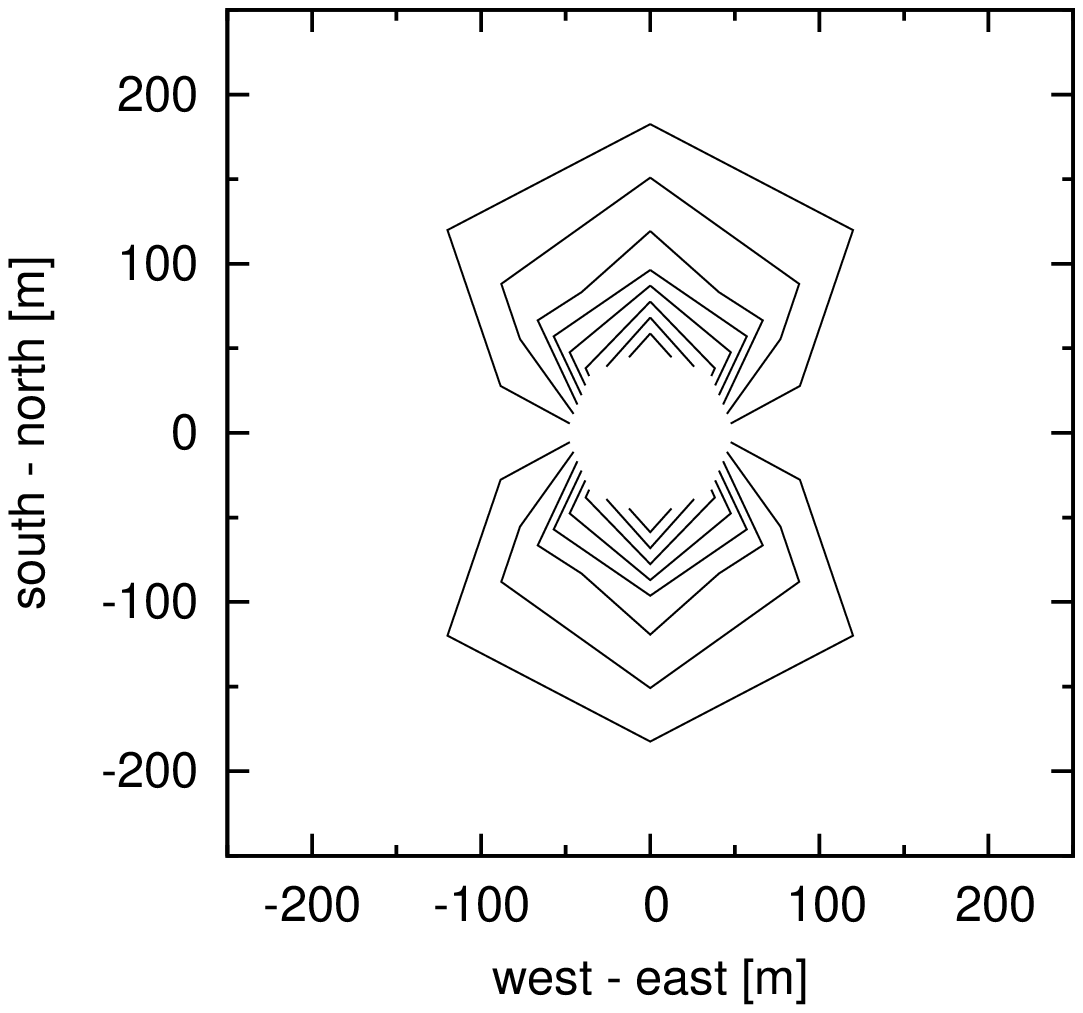}
		\end{minipage}
		\hspace{-0.6cm}
		\begin{minipage}[b]{0.34\linewidth}
		\centering
		\includegraphics[angle = 270 , width = 1.2\textwidth]{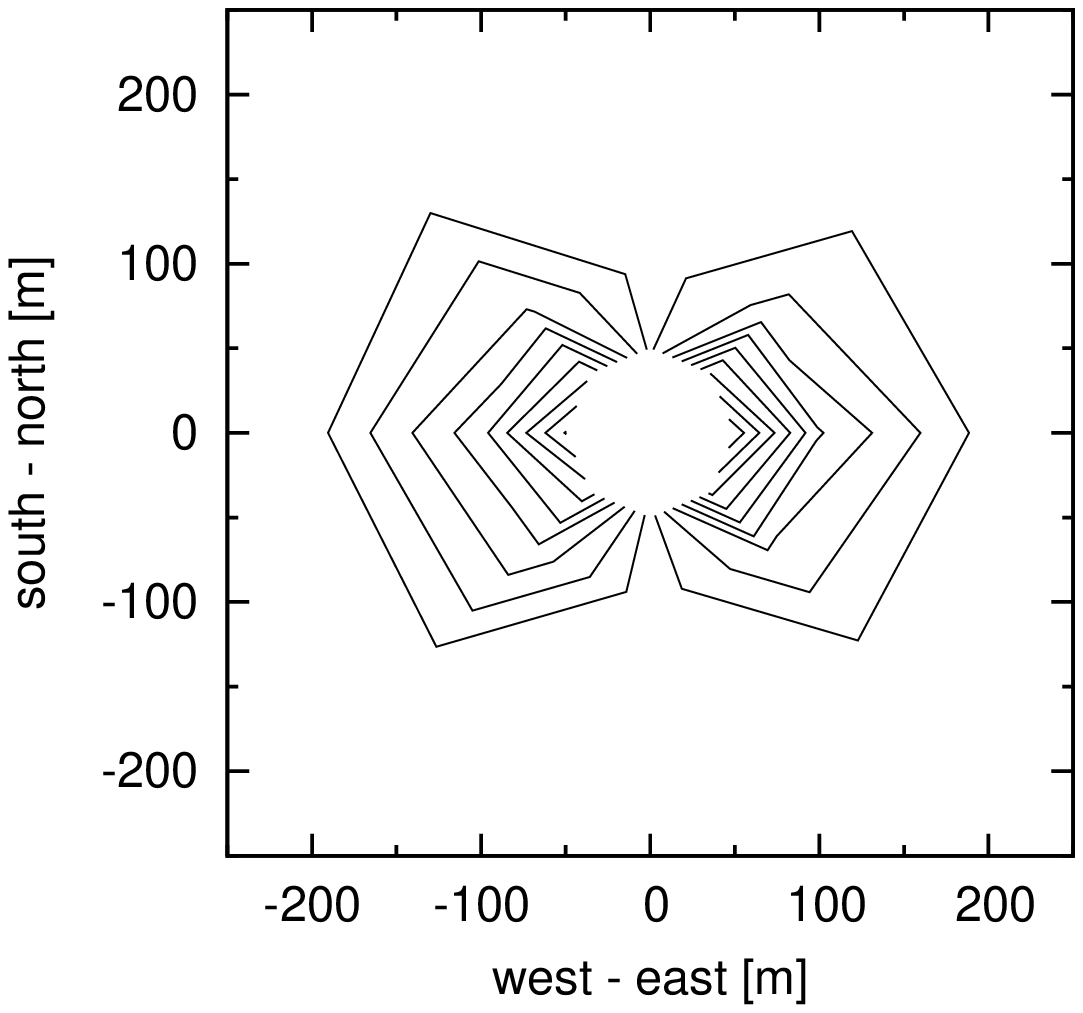}
		\includegraphics[angle = 270 , width = 1.2\textwidth]{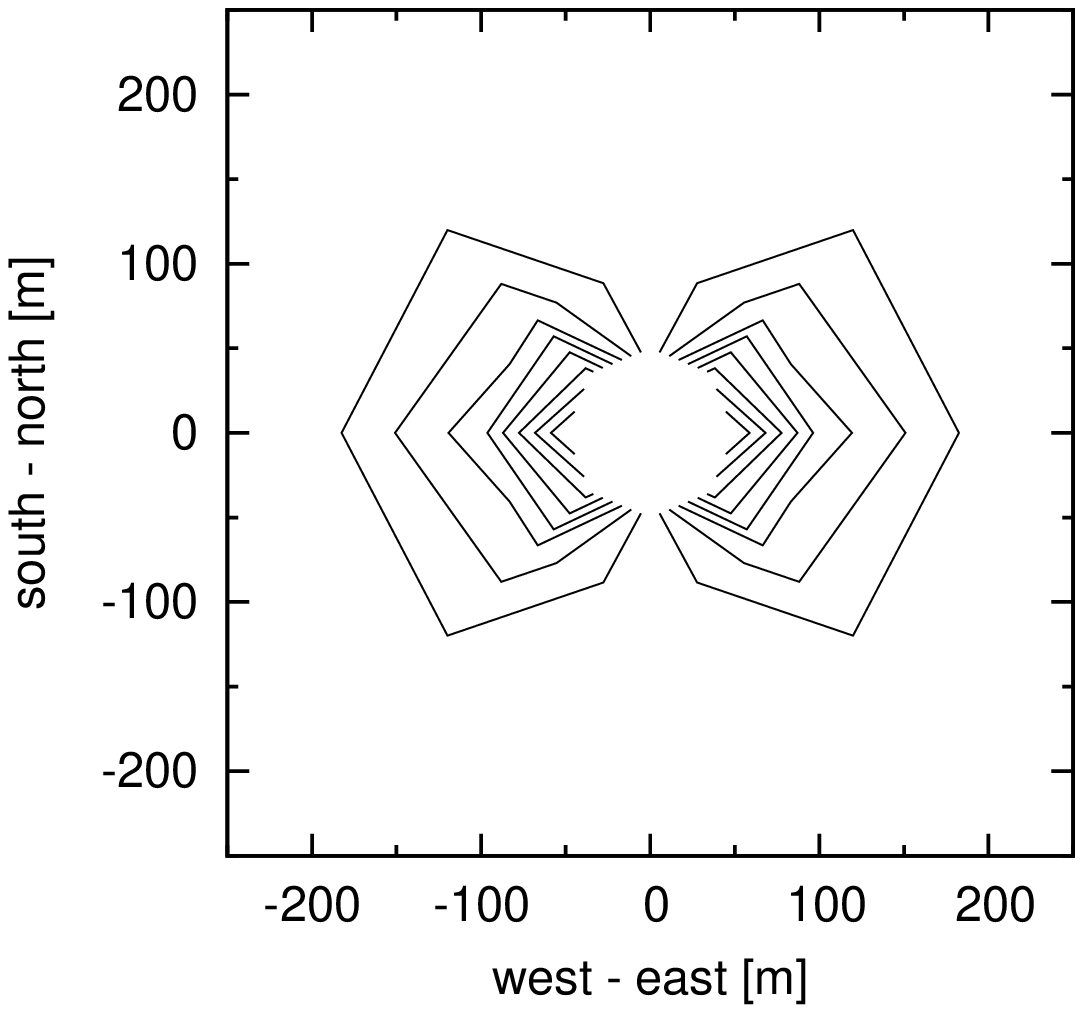}
		\end{minipage}
		\caption[Contour plots of the net charge excess]{Contour plots of the 60\,MHz field strength for the emission from a $10^{17}$\,eV air shower where the magnetic field is switched off completely.
		From left to right:	total field strength, north-south and east-west polarisation component. 
		Contour levels are 0.1\,$\mu$Vm$^{-1}$MHz$^{-1}$ apart. The closest position of the simulated observers to the shower core
		is 50\,m. Upper row: REAS3. Lower row: MGMR} \label{fig:contourNoBfield}
		\end{center}
	\end{figure}	
The two models show also the expected radial polarisation pattern as displayed in the contour plots of figure \ref{fig:contourNoBfield}. The agreement between both models for the two situations shown in this section are direct evidence for the increased understanding of the radio emission mechanism from extensive air showers.

\section{Comparison of MGMR with simplified REAS3}\label{sec:simplified}

Summing up the results of the previous sections, the comparison between REAS3 and MGMR shows an overall agreement within a factor of $\sim 2-3$. Remembering that a few years ago the models predicted even qualitatively different pulse shapes and consequently differing characteristics in the frequency spectra, this agreement is a milestone in the understanding of radio emission from cosmic ray air showers. It should be stressed once more that the models are technically and conceptually very different and completely independent from each other. 

Looking at the details of the results from the comparison between both models, it is obvious that there are still deviations which are too large to be ignored. Especially for the inclined shower studied in section \ref{subs:inclined} the assumption that the shower model is important was made. In several figures, it was shown that the pulse amplitudes predicted by REAS3 are larger than the amplitudes from MGMR, at least for small observer distances to the shower core. The reason for this deviation has to be studied in detail, since also these discrepancies need to be understood. For the future, the aim is to understand the existing differences not within a factor of 2-3, but within less than 10\%. Thus, the influence of the differences in the underlying air shower models has to be studied and it has to be asserted if these are responsible for the deviations. One possibility to study the influence of the air shower models is to simplify the air shower model implemented in REAS3 to get a more similar model to the one used in MGMR and to study the effects on the radio emission. In the following section, this will be done. 

\subsection{The underlying air shower models}\label{subs:modeldiff1}

To adapt REAS3 for comparing the details with MGMR, first the differences of the air shower models 
need to be identified. One major difference is that the spatial distribution of the particles in the shower pancake is different. To mimic the arrival time distribution of MGMR in REAS3, we replaced the longitudinal displacement of the particles in the shower pancake with the $\Gamma-$probability distribution function given in section \ref{subs:MGMR}.

Next, the lateral displacement of the particles in the shower pancake had to be changed. In MGMR, this displacement is not directly considered, but a radiation contribution from a static dipole with a length of 15.0\,m is added to the overall radio signal. Hence, in REAS3, we shifted the electrons by 7.5\,m in the eastern direction of the shower axis and the positrons by 7.5\,m in the western direction and switched off the lateral distribution of the particles in the shower pancake. Further parametrisations used in MGMR have not been considered for this comparison since they influence the predictions less than the modification discussed here. However, they have been briefly discussed at the end of section \ref{sec:method}.
	\begin{figure}[htb]
	\begin{center}
		\begin{minipage}[b]{0.49\linewidth}
		\centering	
		\begin{overpic}[angle = 270, width = 1.0\textwidth]{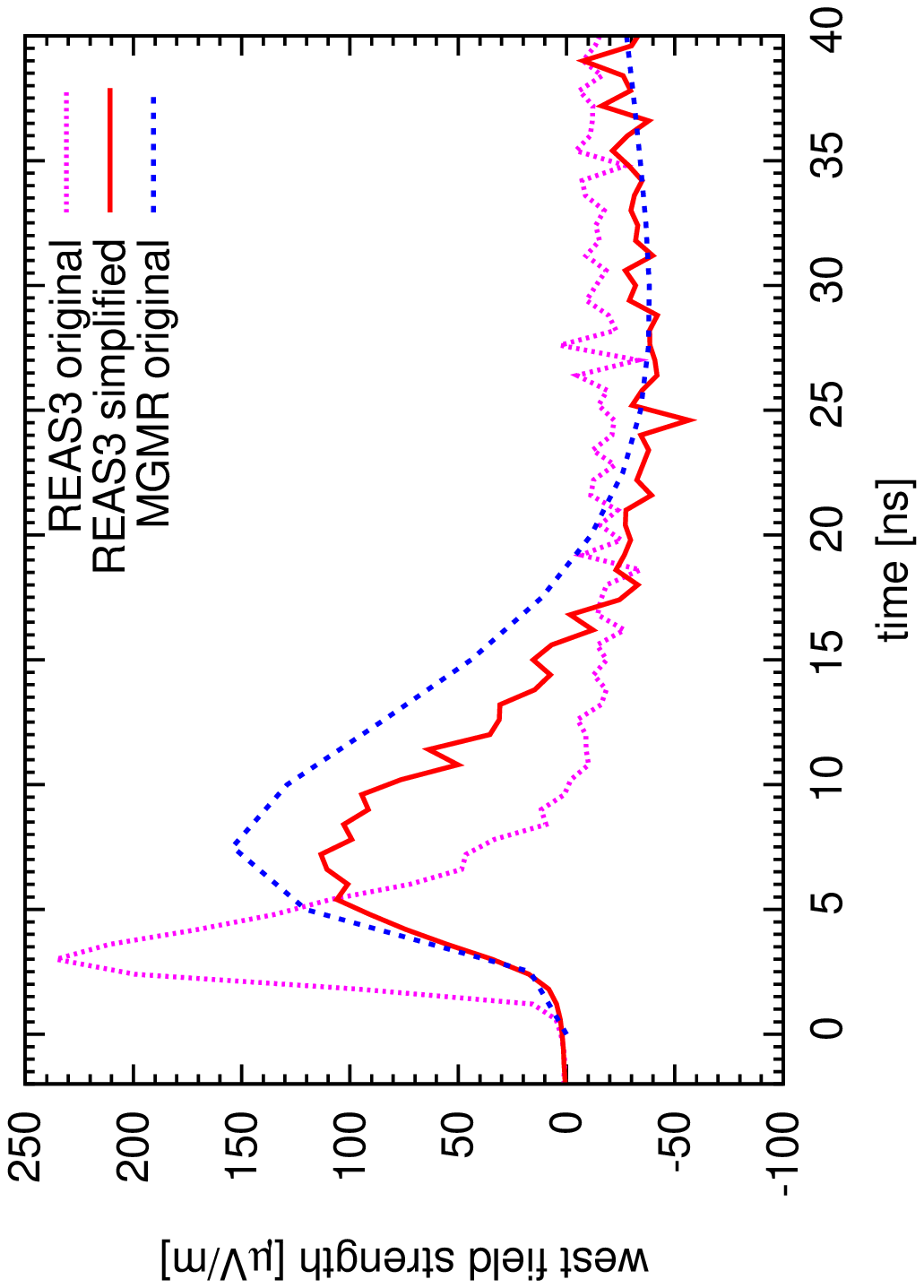}
		\put(60,45){\scriptsize{100\,m - N}}
		\end{overpic}		
		\end{minipage}
		\begin{minipage}[b]{0.49\linewidth}
		\centering
		\begin{overpic}[angle =270, width = 1.0\textwidth]{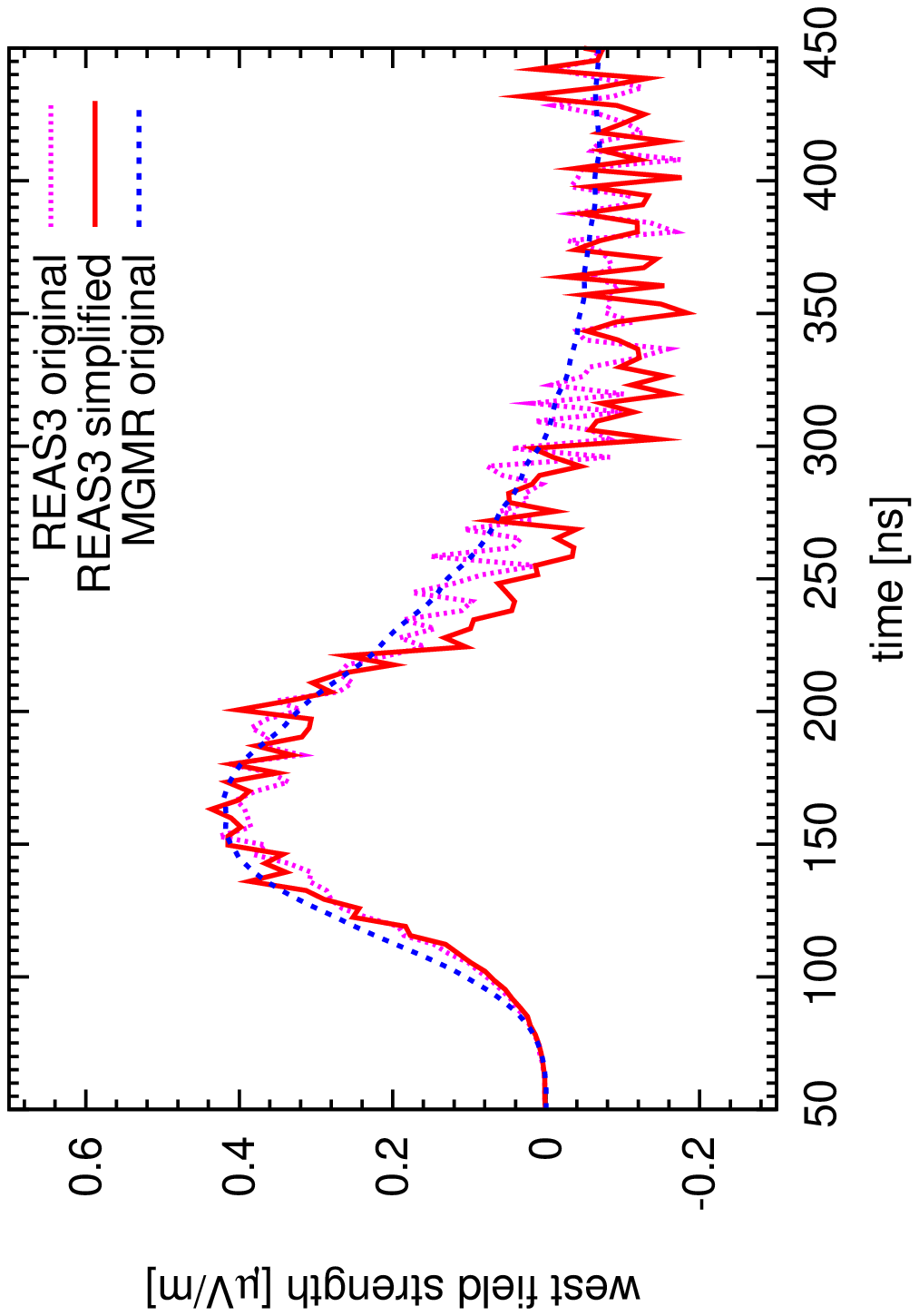}
		\put(60,45){\scriptsize{800\,m - N}}
		\end{overpic}				
		\end{minipage}
		\caption{A comparison between the radio pulses predicted by MGMR and a modified version of REAS3 where the longitudinal displacement of the particles in the shower pancake follows the distribution used in MGMR and the lateral distribution is replaced by a systematic offset of electrons and positrons. The pulses for a $10^{17}$~eV vertical air shower are shown for observers 100\,m (top) and 800\,m (bottom) north of the shower core. }\label{fig:newComparison}
		\end{center}
	\end{figure}
The results when replacing the arrival time distribution and the lateral spread of the particles are displayed in figure \ref{fig:newComparison} for observers at 100\,m north from the shower core and 800\,m north from the shower core of the vertical air shower with primary energy of $10^{17}$\,eV. It is obvious that the modifications influence the results much stronger for small observer distances than for larger observer distances. At 100\,m distance, the pulse amplitude predicted by the modified version of REAS3 deviates by less than $30\%$ from the pulse amplitude predicted by MGMR whereas the original REAS3 pulse amplitude differed by a factor of 1.7. At 800\,m, the modified version of REAS3 obtains the same pulse as MGMR and the original version of REAS3. This underlines the assumption that the details of the air shower model are of little significance for large lateral observer distances. This is reasonable since the time-scales determined by the spatial distributions of the individual particles in the shower pancake dominate at small lateral distances whereas the geometrical time delays resulting from the propagation of the radio emission from the particles in the shower pancake to the observer dominate at larger lateral distances.
	\begin{figure}[htb]
	\begin{center}
		\begin{minipage}[b]{0.49\linewidth}
		\centering	
		\begin{overpic}[angle = 270, width = 1.0\textwidth]{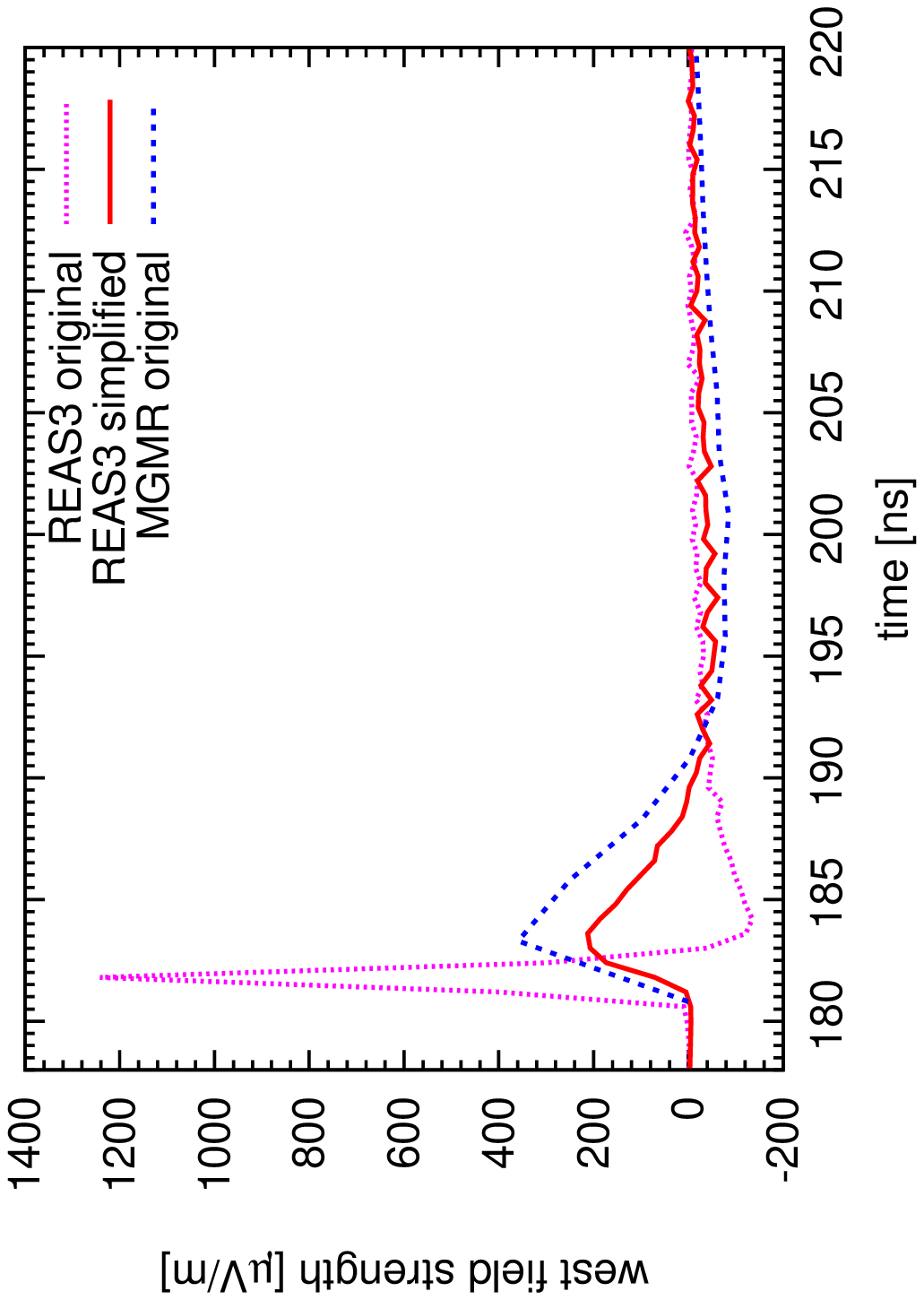}
		\put(60,45){\scriptsize{100\,m - N}}
		\end{overpic}		
		\end{minipage}
		\begin{minipage}[b]{0.49\linewidth}
		\centering
		\begin{overpic}[angle =270, width = 1.0\textwidth]{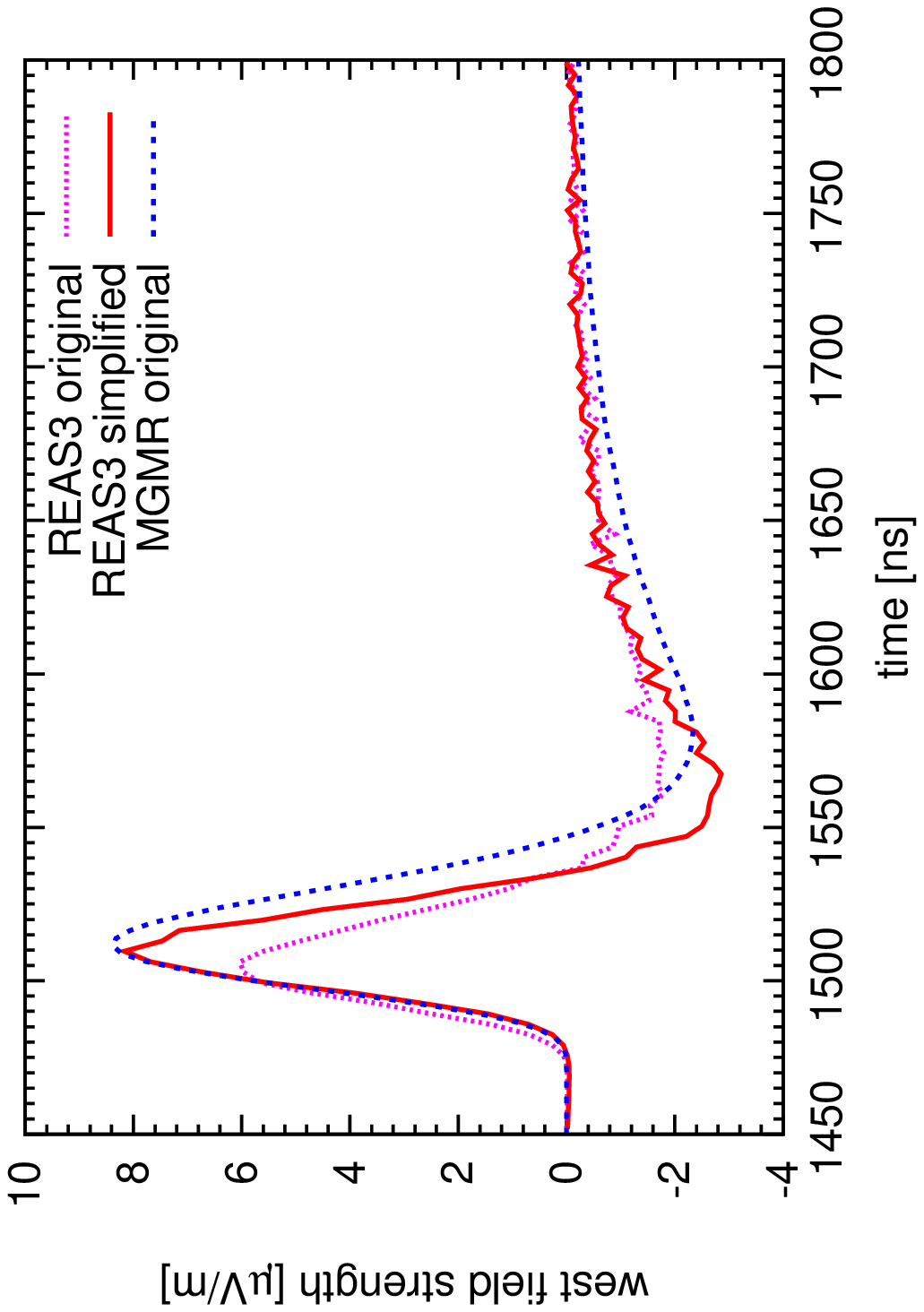}
		\put(60,45){\scriptsize{800\,m - N}}
		\end{overpic}				
		\end{minipage}
		\caption{A comparison between the radio pulses predicted by MGMR and a modified version of REAS3 where the longitudinal displacement of the particles in the shower pancake follows the distribution used in MGMR and the lateral distribution is replaced by a systematic offset of electrons and positrons. The pulses for a $10^{17}$~eV air shower with a zenith angle of $50^{\circ}$ are shown for observers 100\,m (top) and 800\,m (bottom) north of the shower core. }\label{fig:newComparison_incl}
		\end{center}
	\end{figure}

As shown in section \ref{subs:inclined}, the predictions of MGMR and REAS3 differed the most for inclined air showers. We also used the simplified REAS3 to study the influence of the air shower model for the inclined air shower coming from south-east. The results of this comparison are shown in figure \ref{fig:newComparison_incl}. With the original versions of MGMR and REAS3, the predictions differed by a factor of $\sim 3$ for the observer 100\,m north of the shower core. With the simplification of REAS3, the differences became much smaller, i.e. roughly 50\%. Moreover, the arrival times of the pulses were different for REAS3 and MGMR. For the modified version of REAS3 and MGMR, they become similar. As expected, the influence of the details of the air shower model is less at the observer distance of 800\,m but still visible and it leads to a better agreement between the modified version of REAS3 and MGMR.\\ \\
In general, the results of this detailed comparison showed that the differences between REAS3 and MGMR are mostly determined by the different treatment of the distribution of particles in the shower front which is simplified in MGMR by a constant pancake thickness. Since we modified the spatial distribution of particles in the simplified REAS3 version which led to better agreement with MGMR, this gives a strong indication that the pancake thickness in MGMR is overestimated. Hence, in the following we investigated the differences more closely and discuss a proper value for the pancake thickness used in MGMR.

\subsection{Discussion of a realistic pancake thickness in MGMR}\label{subs:results}

As stated in the introduction, the pulse shape is a direct reflection of the important length scales involved in an air shower. In MGMR, the pulse is shown to depend directly on characteristic functions describing the shower evolution, where the most important ones are the shower profile, the drift velocity, the charge-excess in the shower and the particle distributions in the shower front (pancake thickness). 
     \begin{figure}[htb]
     \begin{center}
     	\begin{minipage}[b]{0.65\linewidth}
		\centering
		\includegraphics[angle = 0 , width = 1.0\textwidth]{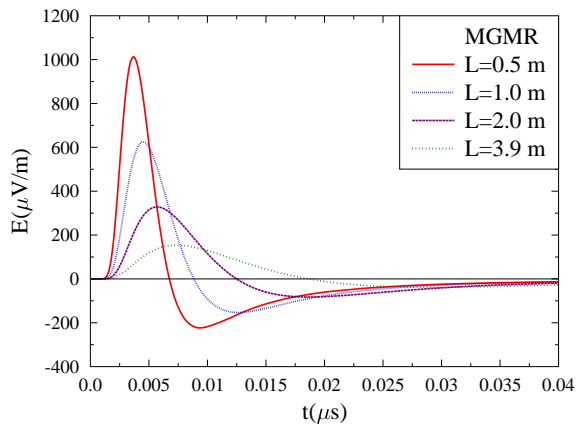}
		\end{minipage}
		\caption[MGMR pancake simulations]{MGMR simulations for different values of the pancake thickness parameter $L$. The simulations are done for a $10^{17}$\,eV vertical shower with the observer positioned $100$\,m to the North of the impact point. The most realistic value for the shower pancake in MGMR for the air showers considered in this paper is $L = 2$\,m (cf. fig \ref{fig:newComparison}).} \label{fig:MGMR_pancake}
		\end{center}
	\end{figure}
The shower profile in MGMR can either be given as a direct input from Monte-Carlo simulations, or, if the previous is not the case, a parametrised profile is used. The drift velocity and the fraction of charge excess act as scale parameters for the geomagnetic and the charge excess radiation, respectively. The pancake thickness and the shower profile functions determine the coherence length and thus the frequency maximum of the pulse power spectrum. At large impact parameters the determining length scale is given by the shower profile and the pancake thickness is less important, while at small impact parameters this is reversed and the pancake thickness becomes the important parameter. With this in mind, we can now also understand why the differences between MGMR and REAS3 are more prominent for small impact parameters. Both simulations use the same shower profile, and hence predict similar pulses for large impact parameters. As shown in this chapter, for small impact parameters the spatial particle distribution in the shower front becomes the important scale parameter. Since REAS3 simulations contain more high frequency components than MGMR in this regime, the logical conclusion would be that the pancake thickness parameter $L$ (Eq.~\ref{eq:longiDisplacement}) in MGMR is overestimated.\\

In the earlier MGMR calculations \cite{Sch08}, the pancake thickness was taken from the measured particle distribution in the shower front at ground, $L=10$\,m,  which is biased towards relatively large distances from the shower core. In the later version~\cite{dVries10} used for this comparison, the average pancake thickness was obtained from Monte-Carlo simulations which resulted in $L=3.9$\,m. Within this comparison between REAS3 and MGMR and with Monte-Carlo simulations~\cite{dvries11}, we have shown that this yields an overestimate since in general the most important part of the electromagnetic pulse is emitted from distances close to the shower axis \cite{LudwigICRC2011} where the pancake thickness is extremely small. In the MGMR implementation for this paper the lateral distribution of the shower particles is ignored, and the effects are taken into account by an effective pancake thickness. Therefore, the value of $L=0.5$\,m as obtained in~\cite{dvries11}, where the lateral distribution is taken into account, is too small and $L=2$\,m is a more realistic value. This also follows from figure~\ref{fig:MGMR_pancake} where we show the influence of the pancake thickness parameter $L$ on the simulated pulse shape. This is done for the $10^{17}$\,eV vertical shower for an observer position $100$\,m north of the shower core and thus has to be compared with figure \ref{fig:newComparison}. Since the relative importance of the real pancake distribution and the lateral distribution depends on the observer geometry the effect is estimated only on average.

\section{Conclusions}\label{sec:conclusion}

In the present article, we discussed a detailed comparison between REAS3 and MGMR. To minimize the influence of differences in the modelling of the atmosphere, this comparison has been carried out for a refractive index of unity. As first result, we achieved an overall agreement within a factor $\sim 2-3$ at short distance, improving with increasing distance from the shower core. This agreement is already a breakthrough in the theoretical understanding of the emission processes of radio emission from cosmic ray air showers because a few years ago, the two models predicted even different pulse shapes and with this deviating frequency spectra. Once again, we want to emphasise that REAS3 and MGMR are technically very different and completely independent from each other. The importance of modelling the air shower and with this the shower front, especially for small observer distances came up when discussing the inclined air shower.

To study the influence of the air shower model on the predicted pulse shapes, the air shower model implemented in REAS3 was modified to get a more similar model to the one used in MGMR and to study the effects on the radio emission. We could verify that the differences in the air shower models are responsible for the observed deviations.

In fact, with the modified version of REAS3 it was possible to reproduce the results derived with MGMR on a 20-50\% level, depending on the shower geometry and the observer position. This indicates that the parametrisation of the thickness of the shower pancake used in MGMR needs to be reconsidered. A more realistic thickness of the pancake in MGMR is achieved with a value of $L = 2$\,m. As this parameter $L$ depends not only on the longitudinal but also on the lateral distribution of the particles in the shower pancake which is not explicitly taken into account in MGMR, it is difficult to make a hard estimate for its value. Furthermore, we showed that for larger observer distances, the details of the underlying air shower models are not very important. Hence in the first part of this comparison, the predictions of MGMR and REAS3 for observers far away from the shower core were already similar.

To conclude, the comparison presented in this article showed the enormous increase of understanding the theory of radio emission from air showers achieved in the recent past.

\section*{Acknowledgements}

This research has been supported by grant number VH-NG-413 of the Helmholtz Association. This work is part of the research program of the ‘Stichting voor Fundamenteel Onderzoek der Materie (FOM)’, which is financially supported by the ‘Nederlandse Organisatie voor Wetenschappelijk Onderzoek (NWO).


\bibliographystyle{model1-num-names}



\end{document}